\documentclass[onecolumn,draftclsnofoot,12pt]{IEEEtran}
\ifCLASSINFOpdf
\else
\fi
\usepackage{amsmath}
\usepackage{amssymb}
\usepackage{amsfonts}
\usepackage{graphicx}
\usepackage{subfigure}
\usepackage{caption} 
\usepackage[ruled,linesnumbered]{algorithm2e}
\usepackage{multirow}
\usepackage{array}
\usepackage{float}
\usepackage{ifpdf}
\usepackage{color}
\usepackage{booktabs}
\usepackage{diagbox}
\usepackage{stfloats}
\usepackage{cite}
\allowdisplaybreaks \allowdisplaybreaks[4]
\newtheorem{theorem}{\bf \emph{Theorem}}
\newtheorem{lemma}{\bf \emph{Lemma}}

\newtheorem{proposition}{\bf \emph{Proposition}}
\begin{document}
\title{Joint Power Allocation and User Association Optimization for IRS-Assisted mmWave Systems}
\author{Dan Zhao, 
	Hancheng Lu, 
	Yazheng Wang,
	Huan Sun,
	Yongqiang Gui
	\thanks{
		This work was supported in part by the National Science Foundation of China (NSFC) (Grants  61831018, 61631017, 91538203).
		
		H. Lu, D. Zhao, Y. Wang, Y. Gui are with the Information Network Laboratory, Department of Electronic Engineering and Information Science, University of Science and Technology of China, Hefei 230027, China (e-mail: hclu@ustc.edu.cn; \{zd2019, wang1997, yongqgui\}@mail.ustc.edu.cn).
		Huan Sun is with the Wireless Technology Laboratory, Huawei Technologies Co.,Ltd., Shanghai, China (e-mail: sunhuan11@huawei.com).
		
	}
}

%
%

\maketitle

\begin{abstract}
Intelligent reflect surface (IRS) is a potential technology to build programmable wireless environment in future communication systems. 
In this paper, we consider an IRS-assisted multi-base station (multi-BS) multi-user millimeter wave (mmWave) downlink communication system, exploiting IRS to extend mmWave signal coverage to blind spots.
Considering the impact of IRS on user association in multi-BS mmWave systems, we formulate a sum rate maximization problem by jointly optimizing passive beamforming at IRS, power allocation and user association. This leads to an intractable non-convex problem, for which to tackle we propose a computationally affordable iterative algorithm, capitalizing on alternating optimization, sequential fractional programming (SFP) and forward-reverse auction (FRA).  
In particular, passive beamforming at IRS is optimized by utilizing the SFP method, power allocation is solved through means of standard convex optimization method, and user association is handled by the network optimization based FRA algorithm.
Simulation results demonstrate that the proposed algorithm can achieve significant performance gains, e.g., it can provide up to $175\%$ higher sum rate compared with the benchmark and $140\%$ higher energy efficiency compared with amplify-and-forward relay. 
\end{abstract}

\begin{IEEEkeywords}
	Intelligent reflect surface, millimeter wave, user association, beamforming, power allocation, auction algorithm.
\end{IEEEkeywords}

\IEEEpeerreviewmaketitle

\section{Introduction}
The fifth generation mobile networks use key technologies such as millimeter wave (mmWave) technology, heterogeneous networks and massive multiple-input multiple-output (MIMO) to cope with increasing demand for higher data rates\cite{2han2019achieving,1zhang2017survey}.
As a promising technology, mmWave communication, with frequency ranging from 30$\sim$300GHz, has the potential to provide gigabits-per-second communication data rates while supporting a wide range of applications \cite{3busari2017millimeter}.
However, due to the high carrier frequency, mmWave suffers more significant path loss compared with the path attenuation over the lower frequency bands. On the other hand, the short wavelength and high directivity cause that mmWave signal are difficult to pass through obstacles, which is highly susceptible to blocking.  
In order to overcome the blockage effect in mmWave systems, thereby improving the coverage of mmWave signal and maintaining robustness of the mmWave systems, 
some research work have made lots of attempts, such as employing relay \cite{niu2019relay, yang2015maximum, zubair2019blockage}, strategically placed reflectors \cite{yiu2009empirical}, the reflections from walls and other surfaces \cite{genc2010robust} to overcome obstacles in mmWave systems.

Recently, intelligent reflect surface (IRS) becomes a new promising technology to improve the coverage enhancement when the line-of-sight (LOS) link is severely blocked by obstacles in mmWave systems \cite{tan2018enabling, wang2019intelligent, 14cao2019intelligent,wang2019joint, zhao2020joint, perovic2020channel, zhang2019reflections}. 
IRS can intelligently reconfigure wireless propagation environment with a large number of passive reflect elements (e.g., low-cost printed dipoles) integrated on a planar surface.
Specifically, each element of IRS receives a superimposed multipath signal from the transmitter and then independently reflects the incident signal by controlling the amplitude and/or phase to achieve passive beamforming (PBF). By smartly turning the phase shifts of the passive elements, the reflected signal can be added coherently at the desired receiver to create effective virtual LOS link while achieving directional signal enhancement, or destructively at non-intended receivers to suppress interference \cite{11wu2019towards}. In addition, compared with relay, such as full-duplex amplify-and-forward (AF) realy, IRS is not equipped with a dedicated power source and an amplifier, which only uses PBF to forward signal, thus the power consumption and hardware costs of IRS are much lower than those of relay \cite{di2020reconfigurable}. 

Many efforts have been dedicated to the research of IRS in mmWave systems \cite{perovic2020channel, wang2019joint, wang2019intelligent, 14cao2019intelligent, zhao2020joint, tan2018enabling, zhang2019reflections}.
A key problem for IRS-assited mmWave systems is to jointly optimize active beamforming at base station (BS) and PBF at IRS to achieve different objectives and improve communication performance, such as optimization of channel capacity \cite{perovic2020channel}, received signal power \cite{wang2019joint, wang2019intelligent} and weighted sum rate (WSR) \cite{14cao2019intelligent}. 
In \cite{perovic2020channel}, the authors firstly presented non-line-of-sight (NLOS) path, then proposed two optimization schemes utilizing IRS to maximize the channel capacity in an indoor MIMO mmWave system. More precisely, one exploits only the customizing capabilities of the IRS elements, while the other jointly optimizes PBF at IRS and transmits active beamforming, which can provide up to $200\%$ channel capacity compared with basic MIMO system.
A problem of maximizing the received signal power was investigated in \cite{wang2019joint, wang2019intelligent} for an multi-IRS assisted multiple input single output (MISO) system, and simulation results demonstrate that both the single IRS and multi-IRS can help to create effective virtual LOS paths and thus substantially improve robustness against blockages in mmWave systems. 
Considering the discrete phase shifts of IRS, the work in \cite{14cao2019intelligent} proposed an alternate optimization algorithm to maximize the WSR of all users in an IRS-assisted multi-user MISO system.
Another key issue is to design an efficient deployment strategy of IRS to establish robust mmWave connections for mobile mmWave systems \cite{tan2018enabling, zhang2019reflections}.
In \cite{tan2018enabling}, the authors developed a two-stage beam-searching algorithm and analyzed the reflect-array deployment to adequately utilize the smart reflectarray, thereby solving the problem of signal blockage in mmWave indoor communications. 
Based on Q-learning and neural networks, a reinforcement learning approach is proposed in \cite{zhang2019reflections} to maximize the downlink transmission capacity, where the location and reflection coefficient of unmanned aerial vehicle-carried intelligent reflector are jointly optimized.

However, existing work on IRS mostly focus on system performance optimization in a single BS scenario \cite{perovic2020channel, wang2019joint, wang2019intelligent, 14cao2019intelligent, tan2018enabling, zhang2019reflections, 11wu2019towards, 19wu2019intelligent, huang2019reconfigurable, kammoun2020asymptotic, cui2019efficient, 15han2019large, wang2020energy} and the multi-BS scenario has not been studied before.
After introducing IRS in a multi-BS scenario, the channel reconfiguration brought by IRS can seriously affect the existing power allocation and user association (UA) algorithms \cite{zhou2017joint, feng2017joint, zuo2016energy, van2016joint, guo2019joint, alizadeh2019load, liu2020user}.
Specifically, the channels involved by IRS are cascaded channels and related to the IRS reflection matrix, which leads to the coupling of the IRS reflection matrix with power allocation and UA. Thus, we have to redesign an effective PBF, power allocation and UA algorithm to enhance the  performance of multi-BS mmWave systems.
Furthermore, because of the erratic nature of mmWave propagation, the mmWave channel varies fastly and is susceptible to blockage effect, causing dramatic block of the communication on a short time scale. All the mentioned mmWave channel characteristics make the optimization of power allocation and UA much more complicated and challenging, and the use of existing power allocation and UA solvers \cite{zhou2017joint, feng2017joint, zuo2016energy, van2016joint, guo2019joint, alizadeh2019load, liu2020user} cannot be applied to actual mmWave systems.
Besides, the reflective elements of IRS usually shift the incident signal with limited discrete phase shifters for the ease of circuit implementation, and this further complicates the joint optimization problem due to the non-convex constraint of IRS.
Nevertheless, lots of work related to IRS assume that each element of IRS is a continuous phase shifter \cite{wang2019joint, 19wu2019intelligent, huang2019reconfigurable, kammoun2020asymptotic}, which is difficult to implement in practice.

To tackle aforementioned issues, in this paper, we study the joint PBF at IRS (i.e., the design of IRS reflection matrix), power allocation and UA optimization problem in IRS-assisted multi-BS mmWave systems. To the best of our knowledge, this is the first attempt to investigate IRS-assisted multi-BS communication. 
It is worth noting that the preliminary design and results have been published in a conference version \cite{zhao2020joint}. Different from the previous work, this paper considers power allocation to further enhance system performance, and provides more comprehensive analyses of the performance of IRS to expand the coverage of mmWave signal. 
By exploiting the benefits from IRS, power allocation and UA, our designed mmWave system can achieve high robust, low complex, and cost-effective communication. 
Our main contributions are summarized as follows:

1) We propose an IRS-assisted multi-BS multi-user mmWave system, where an IRS with discrete phase shifter is deployed to assist downlink MISO transmission. Based on the proposed system,
we jointly take the IRS reflection matrix and transmit power into account so as to formulate the sum rate maximization problem more practically.

2) Given the non-convexity and complexity of the problem, 
we propose an iterative PBF, power allocation and UA (IPPU) algorithm based on alternating optimization technique to solve it.
Particularly, the original problem is decomposed into three subproblems, including the non-convex subproblems of the IRS reflection matrix optimizationn and UA optimization, and the convex subproblem of optimizing power allocation. 

3) It is worth noting that the non-convex subproblems of IRS optimization and UA optimization are very challenging, and we use two effective methods to solve them separately. For the optimization of IRS, we first deduce that the objective function has an upper bound and it satisfies the theoretical requirements of the sequential fractional programming (SFP) method.
Then, we use the monotonic improvement properties of SFP to effectively find the approximate optimal solution of the IRS.
For the optimization of UA, 
we propose a forward-reverse auction (FRA) algorithm to effectively solve this asymmetric multi-assignment problem. 
In particular, we prove that the sum rate of the obtained solution is strictly within 1 of being optimal. 

4) We have further analyzed the performance of the proposed algorithm and compared it with different comparison algorithms. 
Our results show that the proposed algorithm can achieve significant gains in terms of sum rate and energy efficiency (EE).
Specificilly, it can achieve $175\%$ sum rate improvement compared with the benchmark and can also obtain $140\%$ higher energy efficiency improvement compared with AF relay.
Compared with the system assisted by AF relay and the system without IRS, the proposed algorithm has higher coverage and lower outage probability for communication.

The rest of this paper is organized as follows. In Section II, we start with the system and channel model, and then formulate a sum rate maximization problem. In Section III, an IPPU algorithm based on alternating optimization is designed to solve the problem. Simulation results of the proposed algorithm are discussed in Section IV. 
Finally, Section V concludes this paper and gives future research directions.

In this paper, we use italic letters to denote scalars, vectors and matrices are presented by bold-face lower-case and upper-case letters. For any general matrix $\mathbf{A}$, $a_{i,j}$ is the $i$-th row and $j$-th column element. 
$\mathbf{A}^T$, $\mathbf{A}^H$, $\mathbf{A}^{-1}$, $\mathbf{A}^+$ and $\left\|\mathbf{A}\right\|_F$ represent the transpose, Hermitian, inverse, pseudo-inverse and Frobenius norm of $\mathbf{A}$, respectively. 
$\operatorname{Re}\{\cdot\}$, $|\cdot|$, $\operatorname{arg}(\cdot)$ and tr$(\cdot)$ denote the real part, modulus, the angle of a complex vector and the trace of a matrix, respectively.
$\mathbf{A}\otimes \mathbf{B}$ denotes the Kronecker products of $\mathbf{A}$ and $\mathbf{B}$, while vec$(\mathbf{A})$ is a vector stacking all the columns of $\mathbf{A}$. $\mathbf{I}_N$ represents the $N \times N$ identity matrix.
For any general complex vector $\mathbf{x}$, $x_i$ denotes the $i$-th element of $\mathbf{x}$, and $\operatorname{diag}\{\mathbf{x}\}$ denotes a diagonal matrix with each diagonal element being the corresponding element in $\mathbf{x}$. $\left\|\mathbf{x}\right\|_0$ and $\left\|\mathbf{x}\right\|$ denote $\ell_0$ and $\ell_2$ norm of the vector $\mathbf{x}$, respectively. $\mathbb{C}^{x\times y}$ and $\mathbb{R}^{x\times y}$ represent the space of $x \times y$ complex and real number matrices.
Notation $\mathbf{x} \sim \mathcal{CN}(0,\sigma^2)$ means that $\mathbf{x}$ is complex circularly symmetric
Gaussian with zero mean and variance $\sigma^2$, and $E[\mathbf{x}]$ denotes the expected value of $\mathbf{x}$. $j\triangleq \sqrt{-1}$ presents the imaginary unit and the calligraphy upper-case letter $\mathcal{K}$ denotes a set.

\section{System Model and Problem Formulation}
In this section, we describe an IRS-assisted multi-BS downlink multi-user MISO mmWave system model, where each user is served by a BS as illustrated in Fig. \ref{sys}. It should be noted that an IRS is deployed on the surface of the building to assist a severely blocked BS in communication. Then we develop the mmWave channel model to accounts for the channel characteristics of mmWave. This section also describes the problem formulation for the joint design of IRS reflection matrix, transmit power allocation at multi-BS and UA algorithm.
\subsection{System Model}
\begin{figure}
	[t]
	\captionsetup{singlelinecheck = false, justification=justified}
	\captionsetup{font={small}}
	\centering
	\includegraphics[width=0.6\columnwidth]{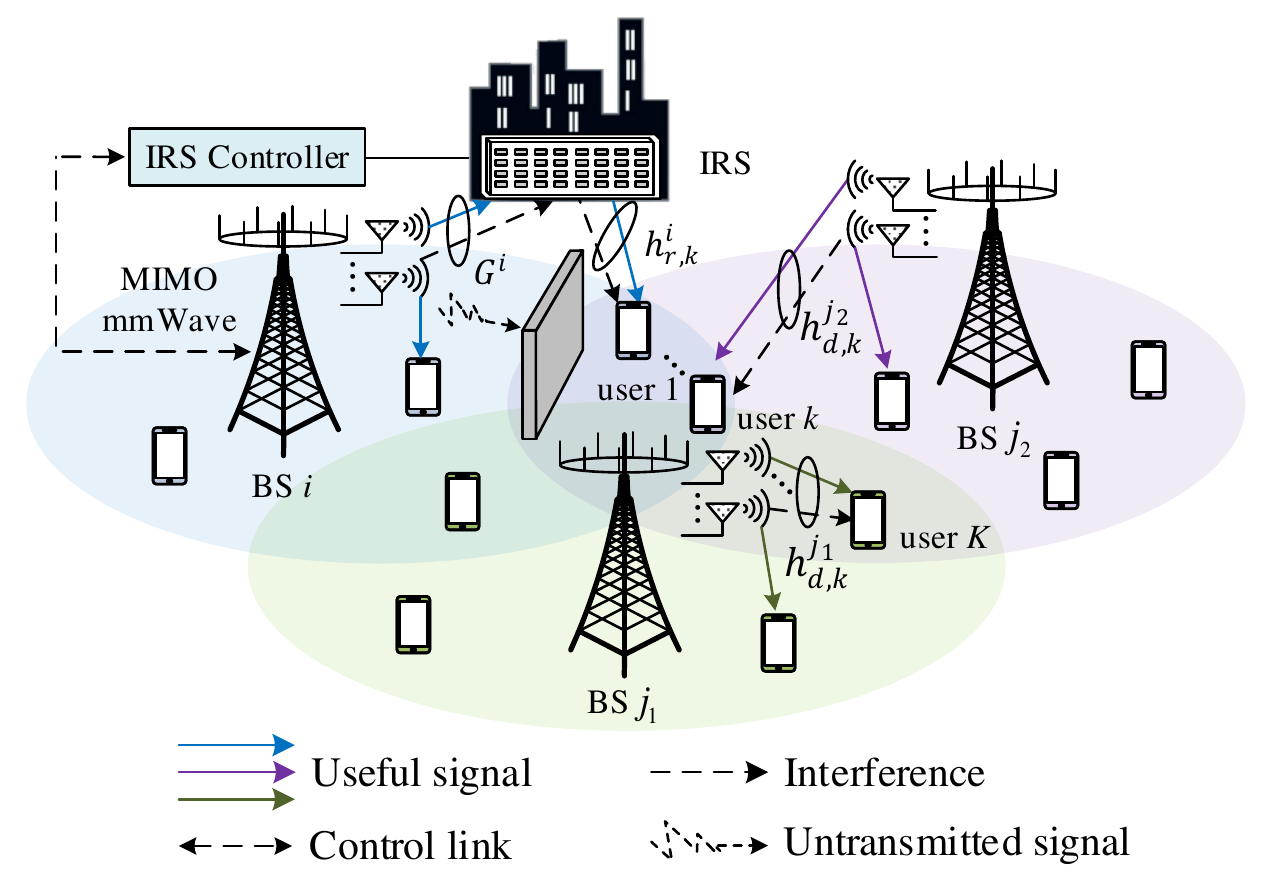}
	\caption{An IRS-assisted multi-BS multi-user mmWave system.}
	\label{sys}
	\vspace{-0.5em}
\end{figure}
As shown in Fig. \ref{sys}, $K$ users are served by $S$ BSs in the designed system, where the set of the user indexes and the BS indexes are denoted as $\mathcal{K}=\{1,2,\cdots,K\}$ and $\mathcal{S}=\{1,2,\cdots,S\}$, respectively. Each BS is equipped with $M>K$ antennas, and each user is equipped with only one antenna. 
Furthermore, we assume that an IRS\footnote{Existing research works \cite{wang2019intelligent, 14cao2019intelligent, wang2019joint} show that IRS-assisted BS communication has strong scalability and signals reflected by multi-IRSs can be superimposed at the receiver to further enhance the received signal power. Therefore, in order to simplify and focus on the impact of the introduction of IRS on the UA and system performance, we only consider one IRS, and multi-IRSs combined with power allocation and UA to enhance mmWave communication will be studied in future work.} with $N$ elements reflects the transmitted signal from the assisted BS $i\in \mathcal{S}$ for its associated users and other BSs $j\in \mathcal{S}, i\neq j$ are not equipped with IRS. 
Equipped with a smart controller, the IRS utilizes passive array to sense the state of the external environment and dynamically adjust the phase shift of each reflecting element.
In downlink transmission, BS can transmit independent data steams to these $K$ users simultaneously.  
We suppose the channel state information (CSI) is perfectly known at BSs and IRS\footnote{Though this is an idealistic assumption, it's still meaningful to study the performance gains brought by IRS and UA for the mmWave system. How to obtain CSI at IRS is out of the scope of this paper, and some related works can be found in \cite{cui2019efficient, 15han2019large}.}, which is the same as \cite{wang2019joint, wang2019intelligent, 14cao2019intelligent, 19wu2019intelligent, huang2019reconfigurable}.

Since the mmWave links are highly susceptible to environmental blockages, we assume that the direct mmWave link between BS $i$ and user $k, k\in\mathcal{K}$ is severely blocked by obstacles, such that BS $i$ can only communicate with user $k$ through the IRS reflection link. In contrast, the channel between BS $j$ and user $k$ is a direct channel.
$\mathbf{G}^i \in \mathbb{C}^{N\times M}, \mathbf{h}^i_{r,k} \in \mathbb{C}^{1\times N}, \mathbf{h}^j_{d,k} \in \mathbb{C}^{1\times M}$ are equivalent channels from the BS $i$ to the IRS, from the IRS to user $k$, and from the BS $j$ to  user $k$, respectively. 
Besides, let $\mathbf{h}^s_k \in \mathbb{C}^{1\times M}$ denotes the channel between BS $s$ and user $k$ and $\mathbf{h}^s_k \in \{\mathbf{h}^i_{r,k}\mathbf{\Phi}\mathbf{G}^i, \mathbf{h}^j_{d,k}\}$, where $\mathbf{h}^i_{r,k}\mathbf{\Phi}\mathbf{G}^i$ represents the concatenation channel between BS $i$ and user $k$. 
The reflection matrix of IRS is denoted as $\mathbf{\Phi} =  \operatorname{diag}\{\beta_1 e^{j\varphi_{1}},\cdots,\beta_N e^{j\varphi_{N}}\}$, where $\beta_n$ and $\varphi_{n}$ represent the amplitude reflection coefficient and the phase shift of the $n$-th element of the IRS, respectively. In practice, each element of the IRS is usually designed to maximize signal reflection \cite{19wu2019intelligent}, and thus we set $\beta_n=1, \forall n$ in the sequel of this paper.

Furthermore, we define the UA matrix $\mathbf{A} \in \mathbb{C}^{S\times K}$ in (\ref{UA}), where $\mathbf{\tilde{a}}_s$ and ${{\left\| {{{\mathbf{\tilde{a}}}}_{s}} \right\|}_{0}}$ denote the UA vector at the BS $s$ and the number of users served by $s$, respectively. And $\mathbf{a}_k={{\left[a_{1,k},a_{2,k},...,a_{S,k} \right]}}^T\in\mathbb{C}^{S\times 1}$ represents the UA vector of user $k$, where $a_{s,k}$ is a binary decision variable (i.e., $a_{s,k} \in \{0, 1\}, \forall s, \forall k$). 
In detail, $a_{s,k}=1$ if user $k$ is associated to BS $s$ and $a_{s,k}=0$ otherwise. 
Since each user is served by only one BS, ${{\left\| {{{\mathbf{a}}}_{k}} \right\|}_{0}}=1$ and we need to determine BS $s$ that actually provides service for user $k$ via UA vector $\mathbf{a}_k$.
Meanwhile, we denote the set of users that BS $s$ can serve as $\mathcal{A}(s)$ and the set of BSs that can serve user $k$ as $\mathcal{B}(k)$. $\mathcal{D}$ is defined as a set of all possible BS-user assignment pairs $(s,k)$ with $k \in \mathcal{A}(s)$.
\begin{equation}\label{UA}
\mathbf{A} = \left[\mathbf{a}_1, \mathbf{a}_2,\cdots , \mathbf{a}_K \right]=\left[
\begin{matrix}
\underbrace{a_{11}, a_{12}, \cdots, a_{1K}}_{\widetilde{\mathbf{a}}_1}\\
\vdots \\
\underbrace{a_{S1}, a_{S2}, \cdots, a_{SK}}_{\widetilde{\mathbf{a}}_S}
\end{matrix}
\right],
\end{equation}

The complex baseband transmitted signal at the BS can be expressed as $\mathbf{x}_s=\sum_{k \in \mathcal{A}(s)}\sqrt{ p^s_k }\mathbf{w}^s_k t^s_k$, where $p^s_k $, $\mathbf{w}^s_k \in \mathbb{C}^{M\times 1}$ and $t^s_k$ denote the transmit power, precoding vector and transmitted data symbol at BS $s$, of $k$-th user, respectively. It is assumed that $t^s_k, \forall s, \forall k,$ are independent variables with zero mean and unit power.
The signal received by user $k$ from BS $s$ can be expressed as
\begin{equation}\label{ysk}
y_{s,k} = \mathbf{h}^s_k \sqrt{ p^s_k }\mathbf{w}^s_k t^s_k + \mathbf{h}^s_k \sum_{j\neq k, j \in \mathcal{A}(s)} \sqrt{p^s_j}\mathbf{w}^s_j t^s_j +n_k,
\end{equation}
where $n_k \sim \mathcal{CN}(0,\sigma^2)$ is complex additive white Gaussian noise with zero mean and variance $\sigma^2$. The power of the transmit signal from the BS $s$ is restricted by the maximum transmit power threshold $P_{max}$
\begin{equation}\label{Ex}
E\left[|\mathbf{x}|^2\right] = \operatorname{tr}(\mathbf{P}_s \mathbf{W}^H_s \mathbf{W}_s) \leq P_{max},
\end{equation}
where $\mathbf{W}_s=\left[{\mathbf{w}^s_{1}},\cdots,{\mathbf{w}^s_K}\right] \in \mathbb{C}^{M\times K}$,  $\mathbf{P}_s=\operatorname{diag}\left[{p}^s_{1},\cdots,{p}^s_K\right] \in \mathbb{R}^{K\times K}$.

In this paper, we suppose that neighboring BSs can be allocated orthogonal frequency band or employ enhanced inter-cell interference coordination techniques \cite{Guoastely2009lte} to eliminate inter-cell interference. Accordingly, the SINR of user $k$ is written as $\gamma_k$ and it can be given by
\begin{equation}\label{SNRk}
\gamma_{s,k} = \frac{ p^s_k \left| \mathbf{h}^s_k \mathbf{w}^s_k \right|^2}{\sum_{j\neq k, j \in \mathcal{A}(s)}\left| p^s_j \mathbf{h}^s_k \mathbf{w}^s_j \right|^2 + \sigma^2}, \forall s, \forall k.
\end{equation}

Then, the achievable rate from BS $s$ to user $k, k \in \mathcal{A}(s)$ is defined as $R_{s,k}$, which can be computed as  
\begin{equation}\label{18}
R_{s, k} = \log_2\left(1 + \frac{p^s_k \left| \mathbf{h}^s_k \mathbf{w}^s_k \right|^2}{ \sum_{j\neq k, j \in \mathcal{A}(s)}\left| p^s_j \mathbf{h}^s_k \mathbf{w}^s_j \right|^2 + \sigma^2}\right).
\end{equation}
\subsection{MmWave Channel Model}
In mmWave systems, the BS-IRS channel $\mathbf{G}^i$ is modeled according to 3D SalehValenzuela channel model \cite{14cao2019intelligent, lin20183} that has been widely used to characterize the mmWave channel:
\begin{equation}\label{Gi}
\mathbf{G}^i = \sqrt{MN}\sum_{g=0}^{G_p}\alpha_g \xi_t \xi_r \mathbf{a}^H_N \left(\theta_{AoA}^{(g)} \right) \mathbf{a}_M \left(\theta_{AoD}^{(g)} \right),
\end{equation}
where $G_p$ denotes the number of NLOS paths, $g = 0$ denotes the LOS path\footnote{A series of studies have shown that mmWave channel normally consist of a limited number of main multipath components, while the scattering paths at sub-6 GHz is generally rich.}, 
$\alpha_g$ indicates the complex gain associated with the $g$-th path, and $\xi_r$ and $\xi_t$ are the receive and transmit antenna gains, respectively.
The parameters $\theta_{AoA}^{(g)}$ and $\theta_{AoD}^{(g)}$ represent the angle of arrival (AoA) and angle of departure (AoD) of the signal reflected by the IRS in the $g$-th path. 
We employ a uniform linear array (ULA) at the IRS \cite{15han2019large}, and thus the array response function of the IRS can be expressed as
\begin{equation}\label{11}
\mathbf{a}_N(\theta_{AoA}) =\frac{1}{\sqrt{N}} \left[1, e^{j2\pi\frac{d}{\lambda} sin\theta_{AoA}},\cdots, e^{j2\pi\frac{d}{\lambda} (N - 1)sin\theta_{AoA}}   \right]\!,\!\!
\end{equation}
where $\lambda$ is the mmWave wavelength and $d$ is the antenna spacing.

Assume that the IRS is coated on the buildings around users which provides a high probability of LOS propagation \cite{14cao2019intelligent, 19wu2019intelligent}. Thus, the channel between the IRS and the $k$-th user can be described as
\begin{equation}\label{hi}
\mathbf{h}^i_{r, k} =\sqrt{N} \alpha_k \xi_t \xi_r \mathbf{a}_N \left(\theta_{AoD} \right),
\end{equation}
where $\alpha_k$ indicates the complex gain and $\theta_{AoD}$ is the AoD of the signal from the IRS to the $k$-th user. Similar to \cite{wang2019joint}, we generate the direct BS-user channels $\mathbf{h}^j_{d,k}$ according to (\ref{hi}).

\subsection{Problem Formulation}
We further define the channel and beamforming matrix of BS $s$ as $\mathbf{H}_s=[\mathbf{h}^{sT}_{1},\cdots, \mathbf{h}^{sT}_K]^T \in \mathbb{C}^{K\times M}$.  
As there are more antennas than users, simple linear processing techniques such as maximum ratio transmission or zero forcing (ZF) are near optimal \cite{van2016joint}. Therefore, we consider using ZF precoding at each BS by setting $\mathbf{W}_{s}^{opt}=|\mathbf{H}_s|^+$ to achieve perfect interference suppression.
In this paper, our goal is to maximize the sum rate of all users by jointly optimizing the PBF at IRS, power allocation and UA in an IRS-assisted mmWave system under reasonable constraints.
Accordingly, the sum rate maximization problem is formulated as
\begin{subequations} \label{p1}
	\begin{align}
	\underset{\mathbf{\Phi},\{\mathbf{P}_s\}_{s=1}^{S},\mathbf{A}} {\mathop{\max}}&~~
	\sum_{k \in \mathcal{K}} \sum_{s \in \mathcal{S}} a_{s,k} \log_2\left(1+ p^s_{k}\sigma^{-2} \right) \label{9a}\\
	s.t.~~& \varphi_n \in \mathcal{F}, \forall n, \label{9b}\\
	& R_k=\sum_{s \in \mathcal{S}} a^s_k \log_2\left(1+ p^s_{k}\sigma^{-2} \right) \geq R_{min}, \forall s, \forall k, \label{9c}\\
	& \operatorname{tr}((\mathbf{H}_s)^+ \mathbf{P}_s (\mathbf{H}_s)^{+H}) \leq P_{max}, \forall s, \label{9d}\\
	& \sum_{s \in \mathcal{B}(k)} a_{s, k} = 1, \forall k, \label{9e}\\
	& \sum_{k \in \mathcal{A}(s)} a_{s, k} \geq 1, \forall s, \label{9f}\\
	& a_{s, k}\in \{0,1\}, \forall (s,k) \in \mathcal{D}, \label{9g}
	\end{align}
\end{subequations}
where $\mathcal{F}=\left\{ 0,\frac{2\pi }{{{2}^{b}}},\frac{2\pi \times 2}{{{2}^{b}}},\cdots,\frac{2\pi \times ({{2}^{b}}-1)}{{{2}^{b}}} \right\}$ is the set of available phase shifts for the IRS, $b$ is the resolution of the phase shifter at IRS, while $R_k$ and $R_{min}$ are the actual rate of user $k$ and the minimum achievable rate of the all users, respectively. 
Also, constraint (\ref{9b}) accounts for the fact that each IRS reflecting element only provides a discrete phase shift, (\ref{9c}) represents the individual  minimum rate constraint of user $k$ (we use $R_{min}$ to characterize users' QoS requirements in this paper), and (\ref{9d}) ensures that the transmit power of BS $s$ is kept below $P_{max}$.
Then, constraint (\ref{9e}) ensures that each user communicates with only one BS. Constraint (\ref{9f}) indicates that each BS can serve multiple users simultaneously and at least one user. 
Given the non-convexity and complexity of the problem, we forward to solve problem (\ref{p1}) efficiently in the sequel.

\section{Proposed Solution}
We cannot assert global optimality due to the non-convex objective function (\ref{9a}) and the non-convex constraint (\ref{9b}).
Thus, a tractable algorithm employing the alternating optimization technique to separately and iteratively solve $\mathbf{\Phi}$, $\{\mathbf{P}_s\}_{s=1}^{S}$ and $\mathbf{A}$ is proposed. 
In particular, for given $\{\mathbf{P}_s\}_{s=1}^{S}$ and $\mathbf{A}$, we optimize the IRS reflection matrix $\mathbf{\Phi}$ based on the SFP method, and then find the optimum $\{\mathbf{P}_s\}_{s=1}^{S}$, when $\mathbf{\Phi}$ and $\mathbf{A}$ are fixed. Next, we solve $\mathbf{A}$ by utilizing the FRA alogrithm. 
Since (\ref{9a}) is upper-bounded on the feasible set, iterating this solution process improves the sum rate value at each iteration and eventually converges in the approximate optimum value of the objective.
Finally, we propose an IPPU algorithm and analyze the complexity of the proposed algorithm.

\subsection{IRS Reflection Matrix Optimization}
When the power allocation matrix $\{\mathbf{P}_s\}_{s=1}^{S}$ and UA matrix $\mathbf{A}$ are fixed, we only need to consider the channel between all users and IRS-assisted BS $i$ in constraint (\ref{9d}). 
Let $\mathbf{H}^i_r=[\mathbf{h}^{iT}_{r,1},\cdots, \mathbf{h}^{iT}_{r,K}]^T \in \mathbb{C}^{K\times N}$ and then problem (\ref{p1}) can be represented as
\begin{subequations} \label{p2}
	\begin{align}
	&\operatorname{find} ~~~ \mathbf{\Phi}  \label{10C}\\
	s.t.~~& \varphi_n \in \mathcal{F}, \forall n, \label{10b}\\
	& \operatorname{tr}((\mathbf{H}^i_{r}\mathbf{\Phi}\mathbf{G}^i)^+ \mathbf{P}_i (\mathbf{H}^i_{r}\mathbf{\Phi}\mathbf{G}^i)^{+H}) \leq P_{max}.\label{10c}
	\end{align}
\end{subequations}
The main challenge in solving problem (\ref{p2}) lies in the fact that the constraint (\ref{10b}) is a non-convex constraint. 
We observe that problem (\ref{p2}) is feasible if and only if the optimal value of the following optimization problem is lower than $P_{max}$
\begin{subequations} \label{p3}
	\begin{align}
	\mathop{\min}_{\mathbf{\Phi}} ~~& 
	f_1(\mathbf{\Phi})=\operatorname{tr}((\mathbf{H}^i_{r}\mathbf{\Phi}\mathbf{G}^i)^+ \mathbf{P}_i (\mathbf{H}^i_{r}\mathbf{\Phi}\mathbf{G}^i)^{+H})  \label{11a}\\
	s.t.~~& \varphi_n \in \mathcal{F}, \forall n. \label{11b}
	\end{align}
\end{subequations}

\begin{theorem} \label{theorem1}
	The optimization problem (\ref{p3}) can be represented as 
	\begin{subequations} \label{13}
		\begin{align}
		\mathop{\min}_{\mathbf{\Phi}} ~~& \mathbf{y}^H  \mathbf{B} \mathbf{y} \label{13a}\\
		s.t.~~& \varphi_n \in \mathcal{F}, \forall n, \label{13b}
		\end{align}
	\end{subequations}
	where
	\begin{subequations}\label{By}
		\begin{align}
		& \mathbf{B}=((\mathbf{\widetilde{H}}^i_r)^{+H}\otimes (\mathbf{G}^i)^+)^H ((\mathbf{\widetilde{H}}^i_r)^{+H}\otimes (\mathbf{G}^i)^+) \in \mathbb{C}^{N^2\times N^2}, \label{B}\\
		& \mathbf{y}=\operatorname{vec}(\mathbf{\Phi}^{-1})\in \mathbb{C}^{N^2\times 1}. \label{y}
		\end{align}
	\end{subequations}
\end{theorem}

\textbf{\emph{Proof}:} The proof of Theorem \ref{theorem1} can be found in the Appendix A of this paper.

As we will show, the objective form in (\ref{13a}) enables us to deal with the non-convex constraint (\ref{13b}), if (\ref{13a}) can be reformulated as a differentiable function.
On this basis, we can use SFP method, also known as Majorization-Minimization method \cite{huang2019reconfigurable, sun2016majorization} to solve problem (\ref{13a}).
The essence of the SFP method is an iterative approach that solves a non-convex problem by solving a sequence of approximate subproblems. If each approximation problem satisfies the following assumptions to the original problem in each iteration, the optimal sequence can decrease monotonically and converges.

1) In each iteration $t$, the feasible sequence $\{\mathbf{y}^{(t)}\}$ maximizes the upper bound of $\mathbf{y}^H  \mathbf{B} \mathbf{y}$.

2) For any $t$-th iteration, the upper bound of the maximum value of the $t$-th iteration must be equal to the true obiective $\mathbf{y}^H  \mathbf{B} \mathbf{y}$, when evaluating at the maximum value calculated in $(t-1)$-th iteration.

The reason is that $\mathbf{y}^H  \mathbf{B} \mathbf{y}$ is the lower-bounded over the feasible set of the problem (\ref{13}).
Nevertheless, the challenge of using the SFP method is to determine a suitable upper bound of $\mathbf{y}^H  \mathbf{B} \mathbf{y}$, which is easier to minimize compared with the original objective (\ref{13a}). Therefore, we provide a convenient upper bound of $\mathbf{y}^H  \mathbf{B} \mathbf{y}$ in the following lemma which fulfills the theoretical requirements of the SFP method.

\begin{lemma}\label{lemma:upbound}
	For any feasible $\mathbf{y}$ and given any feasible point $\mathbf{y}^{(t)}$, a suitable upper bound to employ the SFP method is:
	\begin{equation}\label{15}
	\mathbf{y}^H  \mathbf{B} \mathbf{y} \leq f_2(\mathbf{y}|\mathbf{y}^{(t)}) = \mathbf{y}^H  \mathbf{C} \mathbf{y}+(\mathbf{y}^{(t)})^H (\mathbf{C}-\mathbf{B}) \mathbf{y}^{(t)}\\
	-2\operatorname{Re}\{\mathbf{y}^H(\mathbf{C}-\mathbf{B})\mathbf{y}^{(t)}\},
	\end{equation}
	wherein, $\mathbf{C}=\lambda_{max}\mathbf{I}_{N^2}$, $\lambda_{max}$ is the maximum eigenvalue of $\mathbf{B}$ and the matrix $\mathbf{C}-\mathbf{B}$ is positive semidefinite.
\end{lemma}

	\textbf{\emph{Proof}:} Consider the following inequality:
	\begin{equation}\label{16}
	\|(\mathbf{C}-\mathbf{B})^{1/2} \mathbf{y} -(\mathbf{C}-\mathbf{B})^{1/2} \mathbf{y}^{(t)}\|^2 \geq 0,
	\end{equation}
	then we perform the second-order Taylor expansion of (\ref{16})
	\begin{equation}\label{17}
	\mathbf{y}^H (\mathbf{C}-\mathbf{B}) \mathbf{y}+(\mathbf{y}^{(t)})^H (\mathbf{C}-\mathbf{B}) \mathbf{y}^{(t)}- 2\operatorname{Re}\{\mathbf{y}^H(\mathbf{C}-\mathbf{B})\mathbf{y}^{(t)}\}\geq 0.
	\end{equation}
	At last, the upper bound in (\ref{15}) is obtained by separating and extracting $\mathbf{y}^H  \mathbf{B} \mathbf{y}$, and the proof is complete.

Next, we use the variable $\mathbf{y}$ to rewrite the constraint (\ref{13b}) in detail. Here, it should be noted that (\ref{13b}) only constrains the unit modulus of the discrete phase of the diagonal elements of $\mathbf{\Phi}$, while $\mathbf{y}$ contains the elements of the vectorized $\mathbf{\Phi}$. Therefore, unit modulus of the discrete phase are only allocated to some elements in $\mathbf{y}$ and all other elements must be zero. 
Let $y_i$ be the $i$-th element of $\mathbf{y}, i=1,2,...,N,...,N^2$. For $y_i$ with unit modulus, $y_i=e^{j\theta_i}, \theta_i \in \mathcal{F}$. Then the element $y_i=(i-1)N+i$ must have a discrete phase unit modulus.
From the above, each iteration of the SFP method needs to solve the optimal variable $\mathbf{y}$ through the following minimization problem
\begin{subequations} \label{f2}
	\begin{align}
	\mathop{\min}_{\mathbf{y}} ~~& f_2(\mathbf{y}|\mathbf{y}^{(t)}) \label{18a}\\
	s.t.~~& \theta_i \in \mathcal{F}, \forall i=(n-1)N+n, n=1,2,...,N, \label{18b}\\
	&|y_i|=1, \forall i=(n-1)N+n, n=1,2,...,N, \label{18c}\\
	&|y_i|=0, \forall i\not=(n-1)N+n, n=1,2,...,N, \label{18d}
	\end{align}
\end{subequations}
where
\begin{equation}\label{19}
f_2(\mathbf{y}|\mathbf{y}^{(t)}) = \lambda_{max}\|\mathbf{y}\|^2+(\mathbf{y}^{(t)})^H (\lambda_{max}\mathbf{I}_{N^2}-\mathbf{B}) \mathbf{y}^{(t)} \\
 -2\operatorname{Re}\{\mathbf{y}^H(\lambda_{max}\mathbf{I}_{N^2}-\mathbf{B})\mathbf{y}^{(t)}\}.
\end{equation}
Observing (\ref{18c}) and (\ref{18d}), we can get $\|\mathbf{y}\|^2=N$. By deleting items that are not related to $\mathbf{y}$, problem (\ref{f2}) can be equivalently transformed as
\begin{subequations} \label{20}
	\begin{align}
	\mathop{\max}_{\mathbf{y}} ~~& 2\operatorname{Re}\{\mathbf{y}^H(\lambda_{max}\mathbf{I}_{N^2}-\mathbf{B})\mathbf{y}^{(t)}\} \label{20a}\\
	s.t.~~& (\ref{18b})-(\ref{18d}), \label{20b}
	\end{align}
\end{subequations}
Obviously, the only free variable is the phase $\theta_i$ of $y_i$.
We define $\mathbf{d}=(\lambda_{max}\mathbf{I}_{N^2}-\mathbf{B})\mathbf{y}^{(t)}$,  $d_i$ is the $i$-th element of $\mathbf{d}$, and denote the argument of $d_i$ by $\vartheta_i$. Thus, the optimal phase solution $\theta_i, \forall i=(n-1)N+n, n=1,...,N$ of $\mathbf{y}$ is
\begin{equation}\label{opttheta}
\theta_{i}^{opt}=\arg \underset{\theta_i \in \mathcal{F}}{\mathop{\min }}\,\left| \vartheta_i -\theta_i \right|,
\end{equation}
Hence, for any $\mathbf{y}^{(t)}$, the problem (\ref{f2}) is solved by 
\begin{equation}\label{yopt}
y_i^{opt}=\left\{
\begin{aligned}
&e^{j\theta_{i}^{opt}}, &\forall i=(n-1)N+n, n=1,2,...,N.\\
&0,& \forall i\not =(n-1)N+n, n=1,2,...,N.
\end{aligned}
\right.
\end{equation}

\subsection{Power Allocation Matrix Optimization}
Since constraints (\ref{9b}) and (\ref{9e})-(\ref{9g}) are only determined by IRS reflection matrix $\mathbf{\Phi}$ and UA matrix $\mathbf{A}$, respectively, the problem (\ref{p1}) can be expressed as  
\begin{subequations} \label{23}
	\begin{align}
	\mathop{\max}_{\{\mathbf{P}_s\}_{s=1}^{S}}~~&
	\sum_{k \in \mathcal{K}}  \log_2\left(1+ p^s_{k}\sigma^{-2} \right) \label{23a}\\
	s.t.~~& p^s_{k} \geq \sigma^{2} (2^{R_{min}}-1), \forall s, \forall k, \label{23b}\\
	& \operatorname{tr}((\mathbf{H}_s)^+ \mathbf{P}_s (\mathbf{H}_s)^{+H}) \leq P_{max}, \forall s. \label{23c}
	\end{align}
\end{subequations}
It can be seen that problem (\ref{23}) is convex and can be handled with means of standard convex optimization \cite{jindal2005sum}. Let $\mathcal{P}\triangleq \{ \mathbf{P}_s=\operatorname{diag}\left[{{p}^s_{1}},\cdots,{{p}^s_K}\right]$ : (\ref{23b}) $\&$ (\ref{23c}), $\forall s\}$, we have
\begin{equation}\label{24}
\mathbf{P}_s^{opt}=\arg \underset{\mathbf{P}_s \in \mathcal{P}}{\mathop{\max }}\,\sum_{k \in \mathcal{K}} \log_2\left(1+ p^s_{k}\sigma^{-2} \right), \forall s.
\end{equation}
Furthermore, by analyzing the Karush–Kuhn–Tucker (KKT) optimality conditions of (\ref{24}), the closed-form expression of the solution for $\mathbf{P}_s$ is obtained as
\begin{equation}\label{25}
p_k^{opt}=\mathop{\max}\{0, \varpi \lambda_k-\sigma^{2}\} + \sigma^{2}(2^{R_{min}}-1)\lambda_k^{-1}, \forall k,
\end{equation}
where water level $\varpi$ is the Lagrange multiplier associated to (\ref{23c}) and it follows
\begin{equation}\label{26}
\varpi=\frac{1}{u}\left(P_{max}-\sigma^{2}\left(2^{R_{min}}-2\right)\sum_{k=1}^K \lambda_k^{-1}\right).
\end{equation}
Here, $\lambda_k$ is the $k$-th eigenvalue of $\mathbf{H}_s\mathbf{H}_s^H$ and $u$ is the  number of non-zero eigenvalues $\lambda_k$.

\subsection{UA Matrix Optimization}
In this subsection, we focus on problem (\ref{p1}) to optimize the UA matrix $\mathbf{A}$. To ensure the QoS constraints (\ref{9c}) of all users, we assume that $R_{s, k}=\log_2(1+ p^s_{k}\sigma^{-2}) \geq R_{min} $ holds for all BSs and users.
Considering fixed $\mathbf{\Phi}$ and $\{\mathbf{P}_s\}_{s=1}^{S}$, we pay attention to solving 
the asymmetric multi-assignment problem
\begin{subequations} \label{27}
	\begin{align}
	\mathop{\max}_{\mathbf{A}} ~~& \sum_{s \in \mathcal{S}}\sum_{k \in \mathcal{K}} \frac{R_{s, k}}{R_{min}} a_{s, k} \label{27a}\\
	s.t.~~& \!\!\!\sum_{s \in \mathcal{B}(k)} a_{s, k} = 1, \forall k, \label{27b}\\
	& \!\!\!\sum_{k \in \mathcal{A}(s)} a_{s, k} \geq 1, \forall s, \label{27c}\\
	& \! a_{s, k}\in \{0,1\}, \forall (s,k) \in \mathcal{D}. \label{27d}
	\end{align}
\end{subequations}
Note that $R_{s,k}/R_{min}$ is the benefit of assignment pair $(s, k)$ and the objective function of (\ref{27a}) is the total network benefit.
We define $\mathcal{C}$ as a subset of $\mathcal{D}$, where each BS $s$ is a part of at least one pair $(s, k)\in \mathcal{C}$  and each user $k$ is a part of only one pair $(s, k)\in \mathcal{C}$. 
By setting  $a_{s,k}=1$ if $(s, k)\in \mathcal{C}$ and $a_{s,k}=0$ otherwise, we can obtain a feasible solution of problem (\ref{27}) as well defined as a feasible assignemnt $\mathcal{C}$.
Since the structure of problem (\ref{27}) is consistent with the typical minimum cost flow problem \cite{23bertsekas1998network, 6athanasiou2013auction}, we convert problem (\ref{27}) into a typical minimum cost flow problem by introducing a virtual node $e$ connected to each BS as
\begin{subequations}\label{28}
	\begin{align}
	\mathop{\min}_{\mathbf{A}} ~~& \!\! \sum_{(s, k) \in \mathcal{C}}  \frac{-R_{s, k}}{R_{min}} a_{s, k} \label{28a}\\
	s.t. ~~&\!\!\! \sum_{k \in \mathcal{A}(s)} a_{s, k} - a_{e, s} = 1, \forall s, \label{28b}\\
	& \!\sum_{s \in \mathcal{S}} a_{e, s} = K - S, a_{e,s}\geq 0, \forall s, \label{28c}\\
	& \!\!\!\!\sum_{s \in \mathcal{B}(k)} a_{s, k} = 1, \forall k, a_{s,k}\geq 0, \forall (s, k)\in \mathcal{D}, \label{28d}
	\end{align}
\end{subequations}
where $a_{s,k}$ is extended to include the supernode $e$. 
By using the terminology of network optimization, we declare that $a_{s,k}=1$ means there is one unit flow between $s$ and $k$ and the optimal solution to (\ref{28}) is the same to the initial asymmetric multi-assignment problem (\ref{27}). 
Constraint (\ref{28b}) ensures that the flow supply of each BS $s$ is one unit, and constraint (\ref{28c}) declares that $e$ is the source node with $K-S$ units of flows. The last constraint (\ref{28d}) ensures that each user is served by only one BS.

To proceed further, we utilize the duality theory \cite{23bertsekas1998network} for this minimum cost flow problem (\ref{28}) and  formulate the dual problem
\begin{subequations}\label{29}
	\begin{align}
	\mathop{\min}_{\pi_s, q_k, \mu} ~~& \sum_{s \in \mathcal{S}} \pi_s + \sum_{k \in \mathcal{K} } q_k + (K - S)\mu \label{29a}\\
	s.t. ~~&\pi_s + q_k \geq \frac{R_{s, k}}{R_{min}}, \forall (s, k)\in \mathcal{D}, \label{29b} \\
	&\mu \geq \pi_s, \forall s \label{29c},
	\end{align}
\end{subequations}
where $\pi_s$, $\mu$ and $q_k$ are all Lagrangian multipliers and they are associated with constraint (\ref{28b}), (\ref{28c}) and (\ref{28d}), respectively. 
The parameter $-\pi_s$ represents the price of each BS $s$, $\mu$ stands for the price of source node $e$ and $q_k$ denotes the price of each user $k$.
We can derive the optimal solution of problem (\ref{27}) from the optimal solution of problem (\ref{29}).

Next, we introduce $\epsilon-Complementary \ Slackness$ ($\epsilon-CS$) to solve problem (\ref{29}). Let $\epsilon$ be a positive scalar. Then an assignment $\mathcal{C}$ and a pair $(\pi,p)$ satisfy $\epsilon-CS$ if
\begin{subequations}\label{30}
	\begin{align}
	&\pi_s + q_k \geq \frac{R_{s, k}}{R_{min}} - \epsilon, \forall (s,k) \in \mathcal{D}, \label{30a}\\
	&\pi_s + p_k = R_{s, k}, \forall(s, k) \in \mathcal{C}, \label{30b}\\
	&\pi_s = \mathop{\max}_{l = 1, 2, \cdots, S} \pi_l, s\ \text{has multi-pairs} \ (s, k)\in \mathcal{C}. \label{30c}
	\end{align}
\end{subequations}

\begin{algorithm}[t]
	\fontsize{12pt}{12pt}\selectfont
	\caption{\mbox{FRA based UA Algorithm}} \label{Auction}
	\SetKwInOut{KwIn}{Input}
	\nlset{1}\KwIn{Initial values of $\mathcal{C},(\pi,q), \epsilon$ and $\mu$.}
	\textbf{Ensure:}{$1)\frac{R_{s, k}}{R_{min}}-q_k \geq \max\limits_{l \in \mathcal{A}(s)}\{\frac{R_{s, k}}{R_{min}}-q_l\} - \epsilon, \forall (s, k) \in \mathcal{C};\ $ $  2)\ (\pi,q)\ \text{satisfy}\ \epsilon-CS\ \text{conditions} \ (\ref{30}).$\\}
	\While{there are unassociated BSs}
	{
		BS $s$ is unassociated in $\mathcal{C}$, find the best user $k_s$ that: \mbox{$k_s = \mathop{\arg \max}_{k \in \mathcal{A}(s)} \left\{\frac{R_{s, k}}{R_{min}} - q_k\right\}$, $ \rho_s = \max_{k \in \mathcal{A}(s)} \left\{\frac{R_{s, k}}{R_{min}} - q_k\right\}$}, $\upsilon_s = \max_{k \in \mathcal{A}(s), k \neq k_s} \left\{\frac{R_{s, k}}{R_{min}} - q_k\right\}$;\\
		\If{$k_s$ is the only user in $\mathcal{A}(s)$}
		{
			$\upsilon_s \rightarrow -\infty$;\\
		}
		$ b_{s,k_s} = q_{k_s} + \rho_s - \upsilon_s + \epsilon = R_{s,k_s}/R_{min} - \upsilon_{s} + \epsilon$;\\
		$q_k = \max\limits_{s \in Q(k)} b_{s,k}$, where $Q(k)$ is the set of BSs that user $ k $ received a bid;\\
		Remove any pair $(s', k)\in \mathcal{C}$ and add the pair $(s_k, k)$ to $ \mathcal{C} $ with $ s_k = \mathop{\arg\max}\limits_{s \in Q(k)} b_{s,k} $;\\
		Update $(\pi, q)$ and $\mu = \mathop{\max}_{s = 1, \cdots, S} \pi_s$.\\
	}
	\While{there are unassociated users}
	{
		User $ k $ is unassociated in $ \mathcal{C} $, find the best BS $ s_k $ that: \mbox{$s_k = \mathop{\arg\max}_{s \in \mathcal{B}(k)} \left\{\frac{R_{s, k}}{R_{min}} - \pi_s\right\}$, $ \zeta_k = \max_{s \in \mathcal{B}(k)} \left\{\frac{R_{s, k}}{R_{min}} - \pi_s\right\}$}, $\upsilon_k = \max_{s \in \mathcal{B}(k), s \neq s_k} \left\{\frac{R_{s, k}}{R_{min}}- \pi_s\right\}$;\\
		\If{$s_k$ is the only BS in $\mathcal{B}(k)$}
		{
			$\upsilon_k \rightarrow -\infty$;
		}
		
		$\delta = \min\left\{\mu - \pi_{s_k}, \zeta_k - \upsilon_k + \epsilon \right\}$,
		add $(s_k, k)$ to $ \mathcal{C}$; \\
		Update: $  \pi_{s_k} = \pi_{s_k} + \delta, \ q_k = \zeta_k - \delta $. \\
	}
	
	\KwOut{A feasible optimal assignment $\mathcal{C}$.}
\end{algorithm}

\begin{proposition}\label{pro:upbound}
	Consider a dual variable pair $(\pi, q)$ and let $\mathcal{C}$ be a feasible solution of problem (\ref{29}). Assuming that $\epsilon < 1/S$ and $R_{sk}/R_{min}$ is an integer $\forall {s, k}$, $\mathcal{C}$ is the optimal solution of problem (\ref{29}), if $\epsilon-CS$ conditions (\ref{30}) are satisfied by $\mathcal{C}$ and $(\pi, q)$. 
\end{proposition}

	\textbf{\emph{Proof}:} The proof naturally results from the Proposition 7.7 in \cite{23bertsekas1998network} and Proposition \ref{pro:upbound} in \cite{6athanasiou2013auction}, thus, it will not be elaborated in this paper for briefness.

Based on proposition \ref{pro:upbound}, problem (\ref{29}) can be solved by our proposed FRA based UA algorithm.
As shown in Algorithm \ref{Auction}, it consists of processes of forward and reverse auction. The forward auction aims to associate each BS with one user, and the reverse one is to assign the remaining users to available BSs. 
In the forward auction process, to begin with, we select the unassociated BS $s$ in $\mathcal{C}$ and find the best user $k_s$ which can provides the maximum benefit $\max_{k \in \mathcal{A}(s)}\{\frac{R_{s, k}}{R_{min}} - q_k\}$ among all users of $\mathcal{A}(s)$ (line 4-7). Then, BS $s$ gives a bid $b_{s,k_s}$ for user $k_s$ and the bidding process ends. 
Next, we assign the users who received bid to the corresponding BS $s$ which provides the highest bid and update $k$'s price (line 8-9). 
If user $k$ is initially assigned to another BS $s'$ at the beginning, remove $(s',k)$ and add $(s_k,k)$ to $\mathcal{C}$ (line 10). When all BSs are assigned to one user and satisfy $\epsilon-CS$ conditions, the forward auction is terminated (line 3-12). There are still some unassociated users due to $K>S$.
In the reverse auction process, we assign the rest of the users to available BSs as shown in line 13-20. 
Specifically, select the unallocated user $k$, and we can find the best BS $s_k$ that provides the maximum benefit $\max_{s \in \mathcal{B}(k)} \{R_{s, k}/R_{min} - \pi_s\}$ among all BS of $\mathcal{B}(k)$ (line 14-17). Then, user $k$ decreases its price (i.e., increases the profit $\pi_{s_k}$ for $s_k$) to attract BS $s_k$, and we associate $k$ to $s_k$. After updating the profit of $s_k$ and the price of $k$, respectively, the reverse auction process (line 13-20) is terminated and we can obtain an optimal assignment $\mathcal{C}$ by a finite number of iterations.

Note that $R_{s, k}/R_{min}$ is not the integer required by proposition \ref{pro:upbound} in geneural. Therefore, we first amplify $R_{s, k}/R_{min}$ by a certain factor, and then round to the closest integer value\footnote{In this way, the fractional part of $R_{s, k}/R_{min}$ is much smaller than the integer part of $R_{s, k}/R_{min}$, thus the influence of the optimal solution obtained after rounding on the true optimal value of problem (\ref{27}) can be ignored.} before running Algorithm \ref{Auction}. 

\begin{proposition}\label{opt_c}
	When $\epsilon< 1/S$, Algorithm 1 terminates after a finite number of iterations with an optimal feasible assignment $\mathcal{C}$ of problem (\ref{29}).
\end{proposition}

	\textbf{\emph{Proof}:} First of all, using the theory of auction algorithm directly \cite{23bertsekas1998network}, if the $\epsilon-CS$ condition is satisfied before Algorithm 1 being executed, it is also satisfied at the end of the forward auction. This ensures that each BS is assigned to one user with the maximum benefit.
	Next, in order to prove the optimality and convergence of the improved reverse auction, we need to prove: 1) the reverse auction still satisfies $\epsilon-CS$ conditions after continuous iteration. 2) the reverse auction terminates after a limited number of iterations, and the optimal assignment $\mathcal{C}$ is obtained. We prove the above two points respectively as follows
	
	1) Suppose the profit of BS $s$ before and after iteration are $\pi_s$ and $\pi_s'$, respectively, BS $s^*$ receives a bid from user $k$ and is assigned to user $k$ during the iteration, then $\pi'_{s^*} =R_{s^*, k}/R_{min}-\upsilon_k+ \epsilon$ holds. Next, by substituting the expression of $\upsilon_k$, we can deduce 
	\begin{equation}\label{31}
	\frac{R_{s^*, k}}{R_{min}}-\pi'_{s^*}= \upsilon_k- \epsilon = \max\limits_{s \in \mathcal{B}(k), s \neq s_k} \left\{\frac{R_{s, k}}{R_{min}}- \pi_s\right\}-\epsilon.
	\end{equation}
	Since $\pi'_{s}\geq \pi_{s}$ holds for any $s$, (\ref{31}) can be transformed into
	\begin{equation}\label{32}
	\frac{R_{s^*, k}}{R_{min}}-\pi'_{s^*} \geq \max\limits_{s \in \mathcal{B}(k)} \left\{\frac{R_{s, k}}{R_{min}}- \pi'_s\right\}-\epsilon.
	\end{equation}
	It can be seen that formula (\ref{32}) shows after iterative assignment, all $(s^*,k)\in \mathcal{C}$ satisfy $\epsilon-CS$ conditions. Consider a pair $(s^*,k)$ that belongs to both the assignment before and after iteration. If BS $s^*$ does not receive a bid during the iteration, $\pi'_{s^*}=\pi_{s^*}$, and there is $\pi'_{s}\geq \pi_{s}$ holds for any $s$. Thus $(s^*,k)$ (\ref{32}) holds before and after iteration. In summary, all pairs $(s^*,k)\in \mathcal{C}$ after iteration satisfy $\epsilon-CS$ conditions.
	
	2) Noting that user $k$ is assigned to BS $s_k$ without changing the original association of $s_k$ in each iteration of the reverse auction, BS $s$ will only receive a finite number of bids after meeting $\pi_s=\mu$. 
	Since $\mu$ is the upper bound of the profit that all BSs can reach, at the end of each iteration, the profit $\pi_s$ obtained by $s$ is equal to $\mu$ or at least increased by $\epsilon$. Therefore, if BS $s$ receives infinite bids from users, the profit of $s$ will grow to positive infinity, which contradicts $\mu = \mathop{\max}_{s = 1, \cdots, S} \pi_s$. Thus, the reverse auction process will terminate after a finite number of iterations.

\begin{proposition}\label{se_opt}
	The final assignment $\mathcal{C}$ obtained by Algorithm 1 is within $S\epsilon$ of the optimal assignment benefit of problem (\ref{27}).
\end{proposition}

	\textbf{\emph{Proof}:} Considering the forward auction of Algorithm 1, $\epsilon>0$ is the increment of each bidding. For any assignment $\mathcal{U}=\{(s,k_s )|\ \forall s\}$, its total benefit satisfies
	\begin{equation}\label{33}
	\sum_{s=1}^S \frac{R_{s, k_s}}{R_{min}} \leq \sum_{k=1}^K q_k +
	\sum_{s=1}^S \max\limits_{k}\left\{\frac{R_{s, k}}{R_{min}}- q_k\right\}.
	\end{equation}
	Define $X^*$ as the total benefit that is obtained by the optimal assignment of problem (\ref{27}) and $Y^*$ as the optimal minimum obtained by Algorithm 1 of dual problem (\ref{29}), respectively:
	\begin{subequations}\label{34}
	\begin{align}
	& X^*=\max\limits_{k_s \in \mathcal{A}(s)} \sum_{s\in\mathcal{S}}\frac{R_{s, k_s}}{R_{min}},\ \text{if}\ s\not =m, k_s\not =k_m,\\
	& Y^*=\min\limits_{q_k,k\in\mathcal{K}}\left\{\sum_{k=1}^K q_k+\sum_{s=1}^S\max\limits_{k_s \in \mathcal{A}(s)}\left\{\frac{R_{s, k}}{R_{min}}- q_k\right\}\right\}.
	\end{align}
    \end{subequations}
	Noting that user $k_s$ satisfies the $\epsilon-CS$ conditions, we can get $\frac{R_{s, k_s}}{R_{min}}-q_{k_s}\geq \max\limits_{k \in \mathcal{A}(s)}\{\frac{R_{s, k}}{R_{min}}-q_k\}-\epsilon$. Since the the price of unassigned user is 0, $\sum_{s=1}^S(q_{k_s}+\max\limits_{k_s \in \mathcal{A}(s)}\{\frac{R_{s, k}}{R_{min}}- q_k\}) =
	\sum_{k=1}^K q_k +\sum_{s=1}^S\max\limits_{k_s \in \mathcal{A}(s)}\{\frac{R_{s, k}}{R_{min}}- q_k\} \geq Y^*
	$ is satisfied. Accordingly, it can be deduced that
	\begin{equation}\label{35}
	\begin{split}
	Y^*&\leq\sum_{s=1}^S\left(q_{k_s}+\max\limits_{k_s \in \mathcal{A}(s)}\left\{\frac{R_{s, k}}{R_{min}}- q_k\right\}\right)\\
	&\leq\sum_{s=1}^S \left(\frac{R_{s, k_s}}{R_{min}}+\epsilon\right) \leq X^*+S\epsilon.
	\end{split}
	\end{equation}
    It presents that the total benefit $\sum_{s=1}^S \frac{R_{s, k}}{R_{min}}$ is within $S\epsilon$ of the optimal
   value $X^*$ of problem (\ref{27}). Now consider $\epsilon< 1/S$, then the assignment $\mathcal{C}$ obtained is the optimal within 1 strictly. 
   The reverse auction can be proved similarly and details are omitted.

\subsection{Sum Rate Maximization}
In this subsection, our proposed IPPU algorithm for the initial problem (\ref{p1}) is summarized in Algorithm 2. As stated before, we solve $\mathbf{\Phi}$, $\{\mathbf{P}_s\}_{s=1}^{S}$ and $\mathbf{A}$ by iterating continuously and alternately till reaching a stable optimal rate.
Specifically, since the optimization of $\{\mathbf{P}_s\}_{s=1}^{S}$ and $\mathbf{A}$ will increase the sum rate value $R_{sum}=\sum_{k \in \mathcal{K}} R_k$ after each iteration and the objective is upper bound over the feasible set of (\ref{p1}), the convergence of Algorithm 2 is guaranteed.

\begin{algorithm}[t]
	
	\fontsize{12pt}{12pt}\selectfont
	\caption{\mbox{IPPU Algorithm}} \label{WSR_Max}
	\SetKwInOut{KwIn}{Input}
	\nlset{1}\KwIn{ $\{\mathbf{H}_s\}_{s=1}^{S}$,$K$,$M$,$N$,$b$,$\sigma^2$,$P_{max}$,$R_{min}$,$(\pi, q)$,$\epsilon$,$\mu$, the tolerance $\xi$, iteration number $t,t'$ that are both set to 1 and the upper bound $T_{max}$, $T_{sfp}$; A feasible solution $\mathbf{\Phi}^{(0)}$, $(\{\mathbf{P}_s\}_{s=1}^{S})^{(0)}$, $\mathbf{A}^{(0)}$ of (\ref{p1}). }
	\While{$|R_{sum}^{(t)}-R_{sum}^{(t-1)}|^2>\xi$ and $t\leq T_{max}$}
	{
		\Repeat{$|\mathbf{\Phi}^{(t')}-\mathbf{\Phi}^{(t'-1)}|^2<\xi$ or $t'>T_{sfp}$;}{
			Obtain $\mathbf{B}$ and $\mathbf{y}$ via Eq.(\ref{By});\\
			Calculate the optimal $\mathbf{y}$ as in (\ref{yopt});\\
			$\mathbf{y}^{(t')}$=reshape$(\mathbf{y})$; $t'=t'+1$;
		}
		Update $\mathbf{\Phi}^{(t)}$ via $\mathbf{y}^{(t')}$ with given $(\{\mathbf{P}_s\}_{s=1}^{S})^{(t-1)}$ and $\mathbf{A}^{(t-1)}$;\\
		\If{(\ref{11a}) evaluated at $\mathbf{\Phi}^{(t)}$ is lower than $P_{max}$}
		{
			Update $(\{\mathbf{P}_s\}_{s=1}^{S})^{(t)}$ with given $\mathbf{\Phi}^{(t)}$ and $\mathbf{A}^{(t-1)}$ by using (\ref{25});\\
			\Else{Break and declare infeasibility;}
		}
		Update $\mathbf{A}^{(t)}$ with fixed $\mathbf{\Phi}^{(t)}$ and $(\{\mathbf{P}_s\}_{s=1}^{S})^{(t)}$ by using Algorithm 1;\\
		Set $t=t+1$.
	}
	
	\KwOut{The optimal $\mathbf{\Phi}^{(t)}$, $(\{\mathbf{P}_s\}_{s=1}^{S})^{(t)}$ and $\mathbf{A}^{(t)}$.}
\end{algorithm}

The computational complexity of the proposed algorithm depends on the number of iterations $T_{max}$ to the outermost layer alternation and the complexity required to solve each subproblem.
For Algorithm 2, it can be seen that the complexity of IRS optimization depends on the number of iterations $T_{sfp}$ of the SFP method, multiplied by the amount of operations performed by each iteration. We notice that each iteration of IRS needs to solve a convex problem with $N$ variables. Besides, the transmist power optimization only needs to solve a convex problem with $K$ variables. Furthermore, we know that convex problems have polynomial complexity in the number of optimization variables,  which is at most quartic complexity.

On the other hand, for UA optimization, the total calculation is mainly determined by the value of $\epsilon$ and the maximum absolute value $\Delta=\max\limits_{(s,k)\in \mathcal{D}}\frac{R_{s, k_s}}{R_{min}}-\min\limits_{(s,k)\in \mathcal{D}}\frac{R_{s, k_s}}{R_{min}}$ with $\forall (s, k)\in \mathcal{D}$ in the FRA process. 
In detail, for forward auction process, the total number of bid calculations for BS $s$ is proportional to $\lceil\Delta/\epsilon\rceil|\mathcal{A}(s)|$. So the total running time of forward auction is $\lceil\Delta/\epsilon\rceil\sum_{s=1}^S|\mathcal{A}(s)|=\lceil\Delta/\epsilon\rceil|\mathcal{D}|=\mathcal{O}(\lceil\Delta/\epsilon\rceil SK)$, where $|\mathcal{D}|$ is the number of $(s,k)$ in $\mathcal{D}$. Similarly, the total running time of the reverse auction is $\mathcal{O}(\lceil\Delta/\epsilon\rceil (K-S)S)$. Considering the fact $K>K-S$, the complexity of UA optimization mainly depends on forward auction.
Therefore, the asymptotic complexity of Algorithm 2 can be expressed as
\begin{equation}\label{36}
\mathcal{O}\left(T_{max}\left(T_{sfp}N^z+K^z+\lceil\Delta/\epsilon\rceil SK\right)\right),
\end{equation}
where $1\leq z \leq 4$.

\section{Performance Evaluation}
\subsection{Simulation Scenario}
We consider an IRS-assisted multi-BS multi-user MISO mmWave communication system which operates at 28\ GHz with bandwidth $B=500$ MHz. $K=16$ single-antenna users are randomly distributed in a circular area at $(150 \ m, 50\ m)$ with a radius of $30\ m$. There are three BSs in $(0\ m, 0\ m)$, $(200\ m, 200\ m)$ and $(300\ m, 0\ m)$, respectively, and each of them is equipped with $M=32$ antennas. The IRS comprising $N=32$ passive elements is deployed at a position $(50\ m, 100\ m)$ to assist BS $i$ for communication.
The complex gain $\alpha_k$ is generated according to a
complex Gaussian distribution $\alpha_k \sim \mathcal{CN}(0,10^{-0.1\kappa})$ and $\kappa=\kappa_a+10\kappa_b \log_{10}(d)+\kappa_c$ with $\kappa_a=61.4$, $\kappa_b=2$, $\kappa_c \sim \mathcal{CN}(0,\sigma_c^2)$ and $\sigma_c=5.8$ dB, and $\alpha_g$ is generated according to $\alpha_k$.The LOS path gain $\alpha_0$ is set the same as $\alpha_k$, while the parameters of NLOS path gain are set as
$\kappa_a'=72$, $\kappa_b'=2.92$ and $\sigma_c'=8.7$ dB, and the values of these parameters follow from \cite{mmWavechannel}.
Other required parameters are set as follows: $b=2$, $\sigma^2=-65$ dBm, $P_{max}=30$ dBm, $G_p=5$, $\xi_t=9.82$ dBi, $\xi_r=0$ dBi, $\epsilon=0.2$.

All simulation results are obtained by averaging $10^5$ channels and users positions realizations, and we apply average performance metrics obtained in this way. Specifically, we first generate $10^3$ scenes and randomly set the locations of all users in each scene. Then, we generate $100$ independent channels for each scene to implement Algorithm 2 and optimize the sum rate.

In order to verify the effectiveness of our proposed IPPU algorithm, we consider different four comparison algorithms in vary scenarios:
\begin{itemize}
	\item {\bf AF relay:} We consider a relevant scheme to replace the IRS structure that includes a conventional $N$-antenna AF relay. Different from the IRS reflection matrix $\mathbf{\Phi}$, we assume $N=8$ antennas are used at the AF relay and the diagonal elements of $N\times N$ complex AF matrix $\mathbf{\Gamma}$ are constrained to maximum relay power instead of unit modulus. More details of AF relay can be found in \cite{di2020reconfigurable, huang2019reconfigurable} and omitted here for briefness.
	The power allocation and UA optimization are the same as our proposed algorithm. 
	\item {\bf PBF + UAPC:} We utilize the UA and power control (UAPC) algorithm proposed in \cite{van2016joint} which ensures that all users are treated fairly and the minimum QoS among users is maximized. Meanwhile, the proposed passive beamforming (PBF) is applied at IRS.
	\item {\bf RPBF+NUBA:} The IRS reflection matrix is not optimized, in which the phases are randomly chosen from $\mathcal{F}$. Then, the nearest-based UA (NBUA) \cite{zuo2016energy} algorithm under our proposed maximal transmit power (MP) method is applied at all BSs.
	\item {\bf Without IRS:} Considering the sum rate of all users, each BS uses MP and the traditional received signal strength indicator (RSSI) based UA algorithm to serve users without IRS, which is utilized as a benchmark. Specifically, for the channels between BS $i$ and all users, we generate the BS-user channels with only NLOS components and large path loss to replace BS-IRS and IRS-user channels.
\end{itemize}

\subsection{Convergence Performance and Impact of IRS Parameter}
\begin{figure}
	[tbp]
	\captionsetup{singlelinecheck = false, justification=justified}
	\captionsetup{font={small}}
	\centering
	\includegraphics[width=0.5\columnwidth]{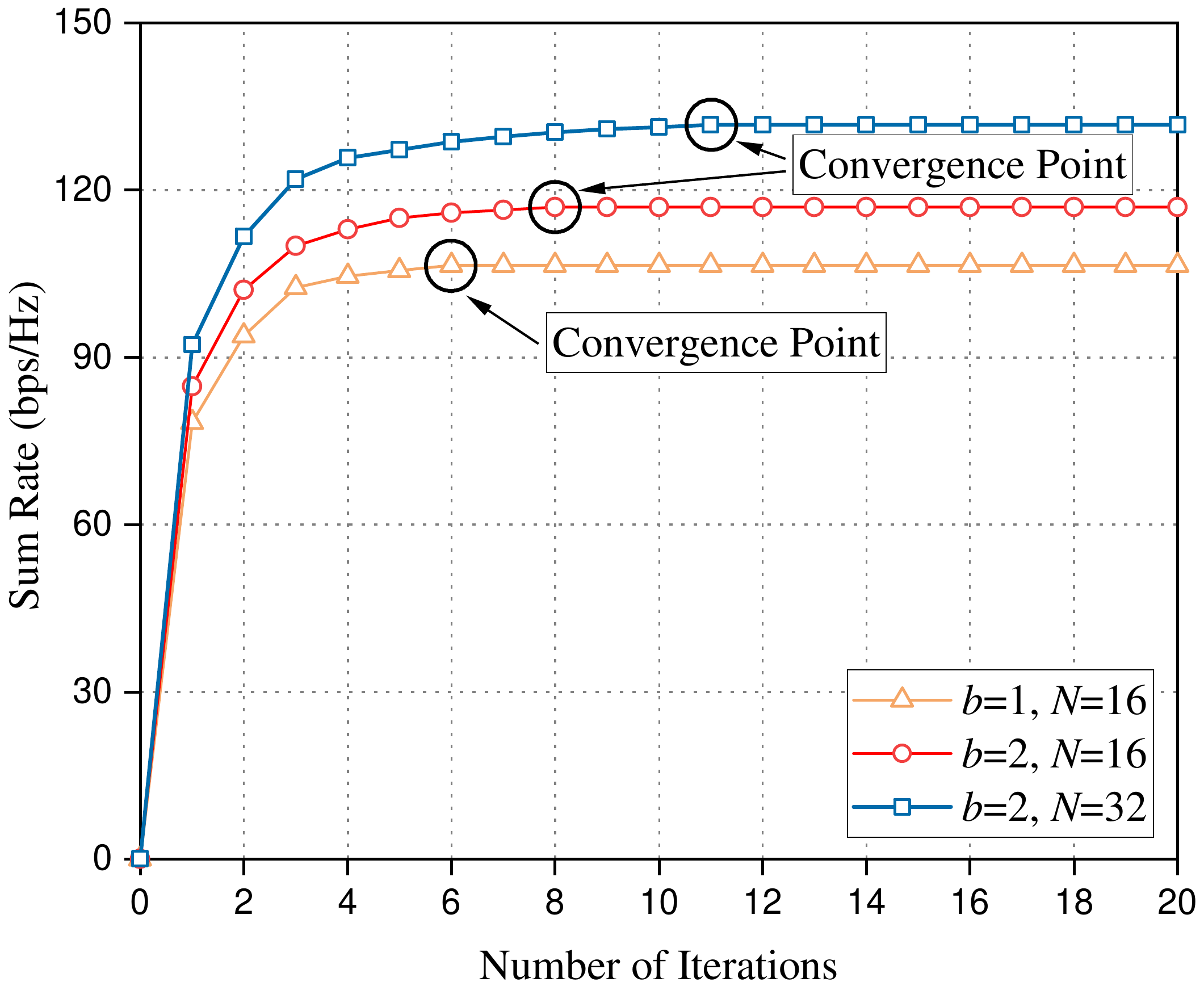}
	\caption{Convergence performance of the proposed algorithm.}
	\label{Iteration}
	\vspace{-1.5em}
\end{figure}

Fig. \ref{Iteration} presents the convergence performance of the proposed algorithm under different number of reflect elements and phase resolutions at IRS. 
It can be seen that, the convergence of our proposed algorithm is confirmed in multiple simulated cases and all these three curves converge to stable solutions after no more than $11$ iterations. When $b=1$ and $N=16$, our proposed algorithm converges after about $6$ iterations. Meanwhile, we observe that the curve with the minimum $b$ and $N$ converges the fastest. With the increase of $b$ and $N$, the convergence speed becomes slower but obtains significant improvement on sum rate. 
On the other hand, when $b$ is equal, the curve with more reflect elements $N$ can achieve higher average sum rate. This is because more signals can be reflected and more effective IRS reflection matrix can be obtained, when there are more reflect elements at IRS. Similarly, utilizing higher phase resolution, IRS can steer the incident signals with finer direction, and thus the curve with higher $b$ can achieve more performance gains.
\vspace{-0.5em}
\subsection{IRS versus AF Relay Performance Comparison}
\begin{figure*}
	[t]
	\captionsetup{singlelinecheck = false, justification=justified}
	\captionsetup{font={small}}
	\centering
	\subfigure[Sum Rate versus $P_{max}$]{
		\label{SEvsP} 
		\includegraphics[angle=0, width=2.85in,height=2.29in]{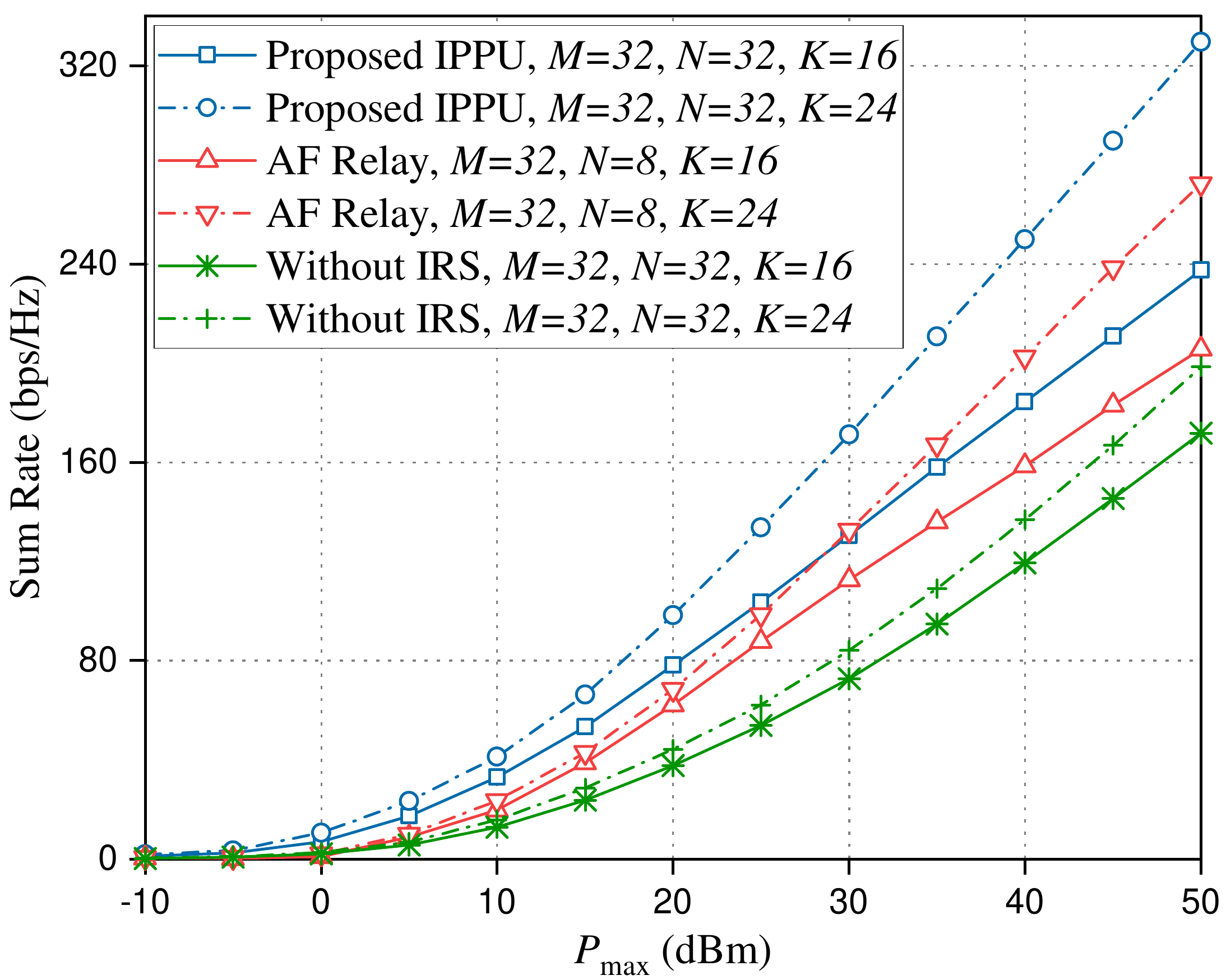}}\hspace{3.0em}
	\subfigure[EE versus $P_{max}$]{
		\label{EEvsP} 
		\includegraphics[angle=0, width=2.83in,height=2.3in]{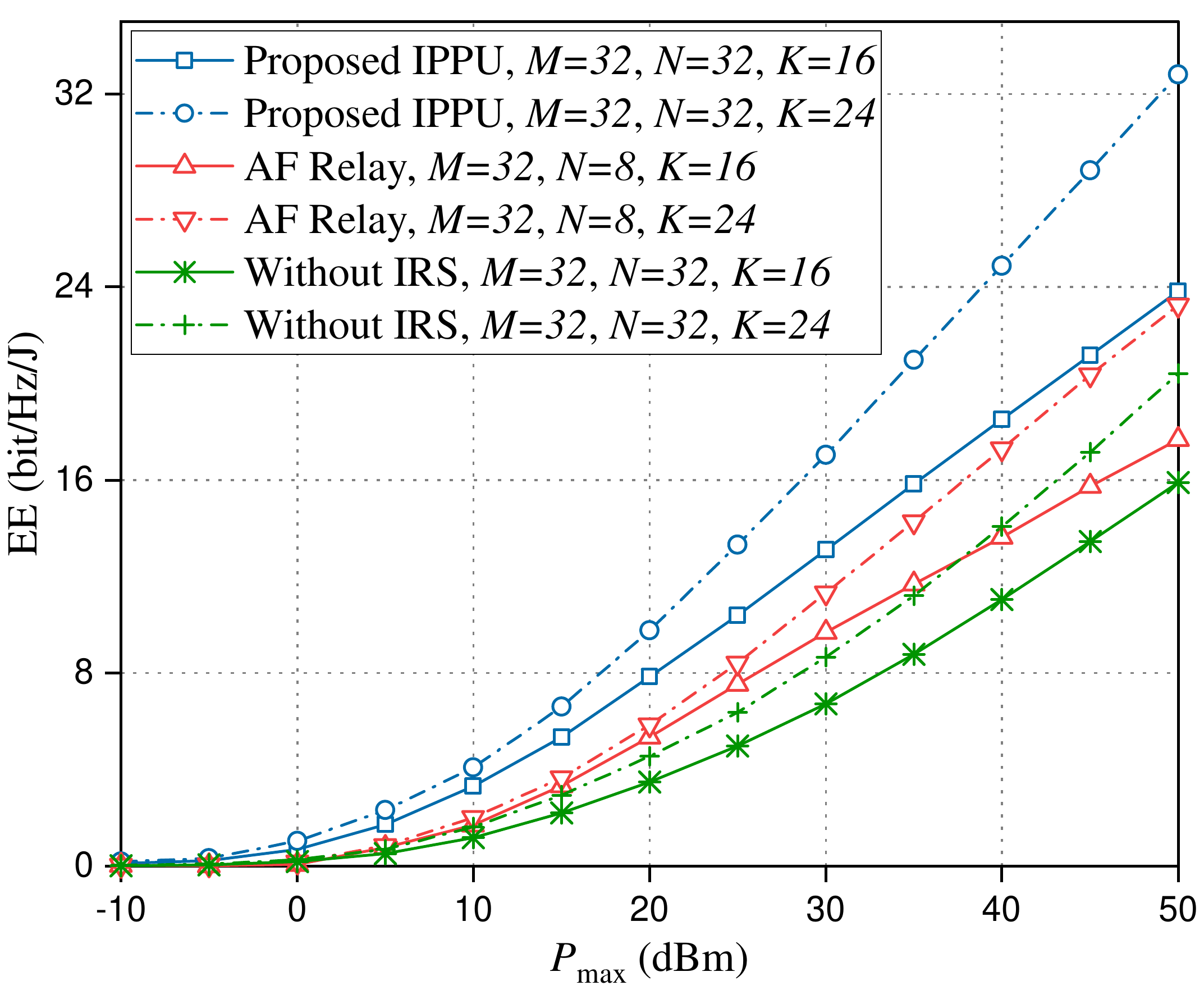}}
	\caption{The performance versus maximum transmit power $P_{max}$ for $R_{min}=0\ $bps/Hz.}
	\label{SEandEEvsP} 
	\vspace{-1.5em}
\end{figure*}
In this subsection, we compare the performance of achievable sum rate and EE with AF relay and benchmark schemes. 
The EE of the considered system enhanced by IRS can be describe as
\begin{equation}\label{ee}
EE=\frac{R_{sum}}{\eta \sum_{(s,k)\in \mathcal{C}} p_k^s + SP_{BS} + KP_u+NP_n},
\end{equation}
where $\eta$ is the circuit dissipated power coefficients, and $P_{BS}$, $P_u$ and $NP_n$ are the circuit dissipated power at BS, user and the IRS, respectively. We ignore $NP_n$ for the benchmark scheme. In the contrast, for AF relay system, $NP_n$ is replaced by $\eta_{AF}P_{AF}+NP_r$ which represents the sum of power consumed by AF realy. Meanwhile, we set $NP_n=0$ for the system without IRS.
Further, we adopt the typical values $\eta=1.2$, $P_{BS}=5$ dBW,  $P_{u}=10$ dBm, $P_n=10$ dBm, while $\eta_{AF}=1.2$, $P_{AF}=2$ dBW and $P_r=10$ dBm\cite{huang2019reconfigurable}.

The sum rate and EE performances under different settings of $P_{max}$ in dBm with $R_{min}=0\ $bps/Hz\footnote{$R_{min}=0$ is equivalent to solving the initial problem (\ref{p1}) without considering the users' minimum rate constraints. Correspondingly, when solving the problem (\ref{27}), $R_{min}$ is set as a very small value to satisfy the use conditions of Algorithm 1.} are illustrated in Fig. \ref{SEandEEvsP}. 
It can be seen from Fig. \ref{SEvsP} that the sum rate performance of the IRS-assisted system outperforms other two schemes with vary settings. 
This is an obvious and predictable result mainly due to strong PBF gain brought by IRS.
For AF relay, it is an active terminal with a dedicated transmitting amplifier circuit and is not restricted by the unit modulus that IRS has. Considering that the number of antennas of AF relay in practical applications is smaller than the number of antennas at BS, the reflection path gain brought by AF relay is still smaller than IRS. 
At the same time, we observe that the performance gap between IRS and AF relay becomes bigger as $P_{max}$ increases. The main reason is that with the increase of $P_{max}$, the relay transmist power $P_{r,max}$ and sum rate become more and more irrelevant and both of them are actually affected by the BS transmist power. 
Specifically, when $P_{max}=50$ dBm and $K=24$, our proposed algorithm can reach up to $120\%$ sum rate compared to AF relay and obtain $165\%$ sum rate compared to the benchmark.
It is also shown that the curve with $K=24$ achieves higher sum rate than that of the curve with $K=16$. This is because, as the number of users increases, the spectrum resource utilization rate and power efficiency of the system increase as well.
Similar results can be observed in Fig. \ref{EEvsP}.

Fig. \ref{EEvsP} presents the EE performance trend under the same settings as Fig. \ref{SEvsP}. In this simulation, our proposed algorithm can obtain the best EE, where the EE of our algorithm is $140\%$ higher than that of relay when $P_{max}=50$ dBm and $K=16$.
This is because the IRS-assisted system has strong PBF gain and lower power consumption, while the power consumption of relay is much greater.
What's more, with increasing transmit power, the EE of all these algorithms increases, which is very similar to the trend of Fig. \ref{SEvsP}.
Due to the channel characteristics of the mmWave, the power allocated by BS to each user is greatly affected by mmWave channel, which leads to a small change in the power allocated by each user within a large maximum power variation range. In other words, the performance of EE is mainly affected by the sum rate rather than the transmit power.
However, we can foresee that the system performance cannot be infinitely improved by increasing transmit power in the actual mmWave system, since the excess BS transmit power is actually not used.
In summary, our proposed algorithm can achieve significant EE gains within an acceptable sum rate range.

\subsection{Impact of BS Parameter and User Parameter}
\begin{figure}
	[t]
	\centering
	\captionsetup{font={small}}
	\captionsetup{singlelinecheck = false, justification=justified}
	\begin{minipage}[t]{0.451\linewidth}
		\centerline{\includegraphics[angle=0, width=2.85in,height=2.29in]{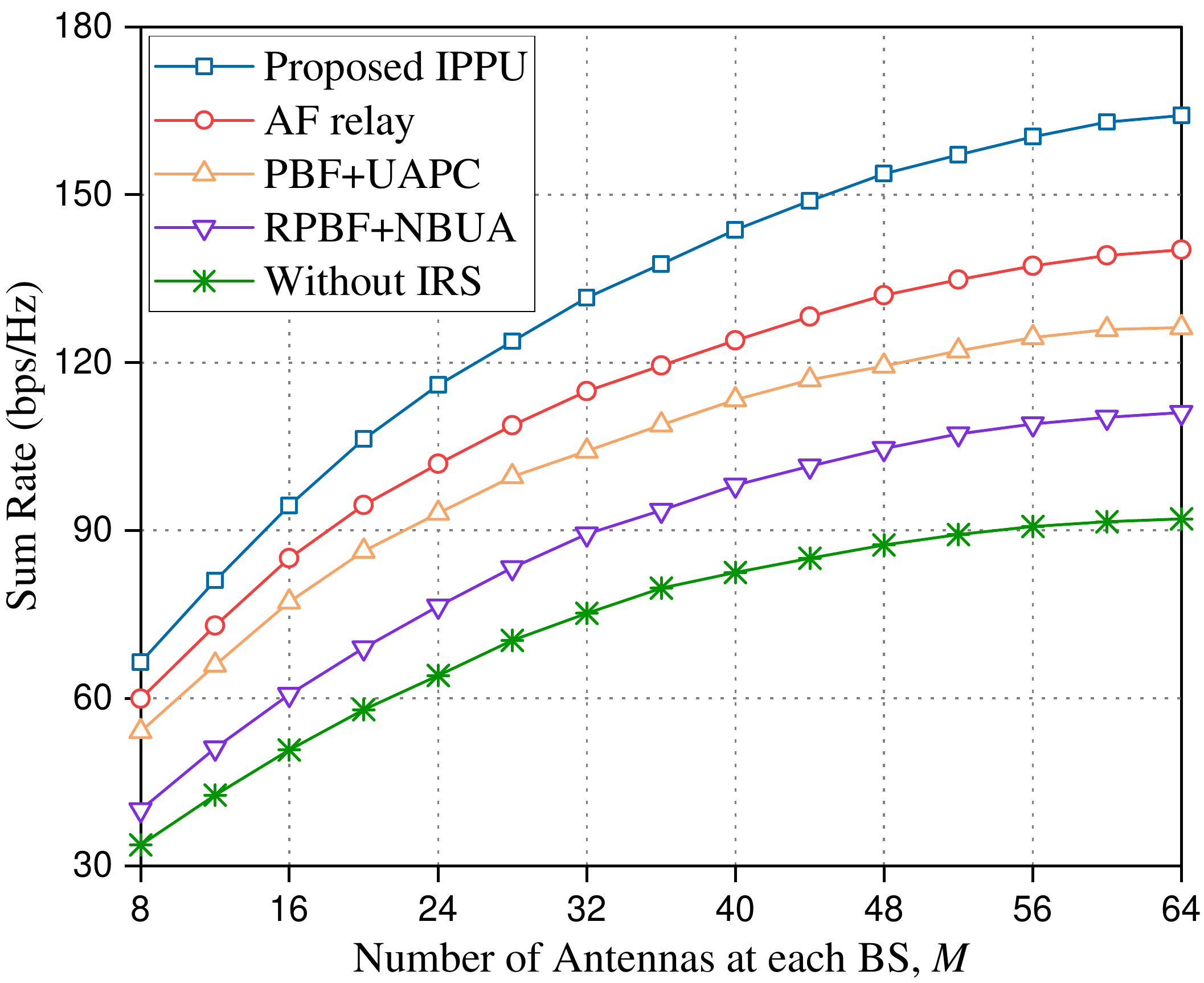}}
		\caption{Sum Rate versus $M$ with $K=16$, $N=32$.}
		\label{M}
	\end{minipage}  \hspace{2.0em}
	\begin{minipage}[t]{0.451\linewidth}
		\centerline{\includegraphics[angle=0, width=2.87in,height=2.28in]{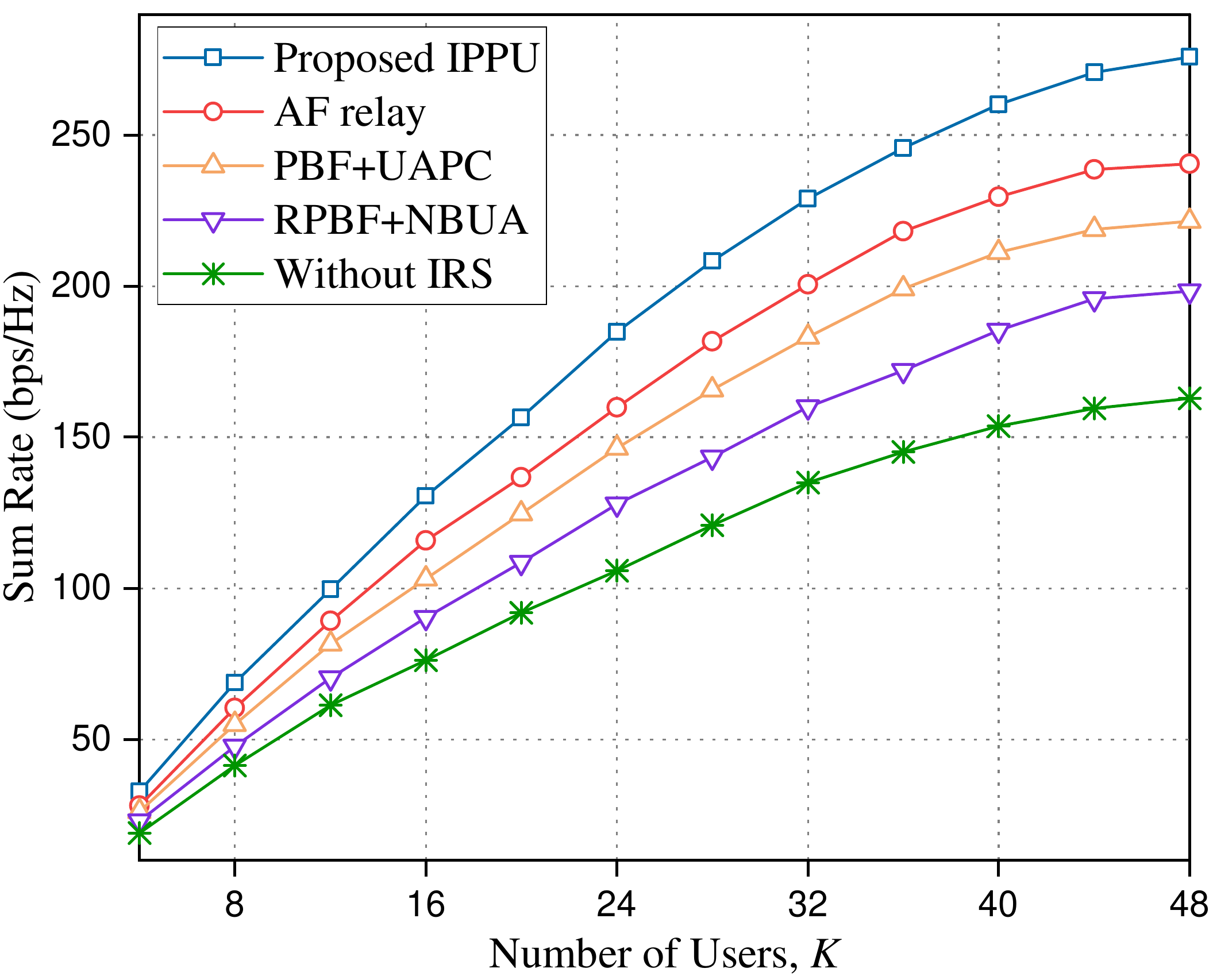}}
		\caption{Sum Rate versus $K$ with $M=32$, $N=32$.}
		\label{K}
	\end{minipage}
	\vspace{-1.5em}
\end{figure}

\begin{figure}
	[t]
	\captionsetup{font={small}}
	\captionsetup{singlelinecheck = false, justification=justified}
	\centering
	\includegraphics[width=0.5\columnwidth]{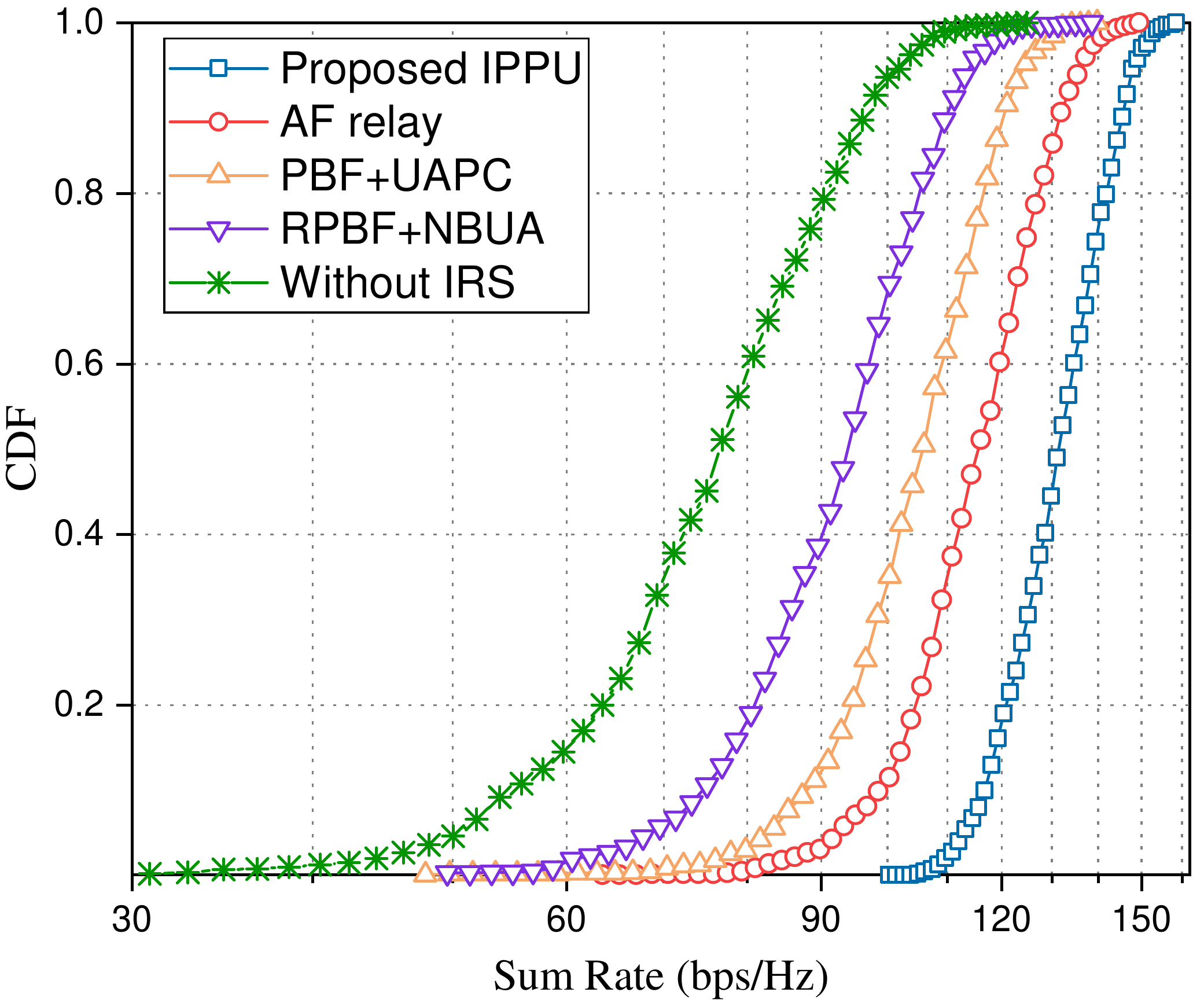}
	\caption{CDFs of sum rate for four comparison algorithms with $K=16$, $M=32$ and $N=32$.}
	\label{CDF}
	\vspace{-1.5em}
\end{figure}

In Fig. \ref{M}, we evaluate the achievable sum rate versus the number of antennas at each BS, where $R_{min}$ is set to 2 bps/Hz, the following four figures (i.e., Fig. \ref{M} - Fig. \ref{Cover_R}) are also the same setting. 
When the number of antennas increases, the sum rate increases and the proposed joint optimization IPPU algorithm gets the maximum sum rate compared with other three algorithms. The reason is that, as the $M$ increases, the number of antennas available for beamforming increases as well, which implies active beamforming becomes more efficient and results in the achievable rate of users to be improved.  
Besides, the proposed algorithm achieves significant gain in sum rate due to better PBF, power allocation and UA gains, which validates the effectiveness of our joint optimization. 
In particular, when $M=32$, the achievable sum rate of the IPPU algorithm is as high as $175\%$ of the benchmark rate, and this conclusion can also correspond to the rate with $K=16, P_{max}=30$ dBm in Fig. \ref{SEvsP}.

Fig. \ref{K} shows the impact of the number of users $K$ on achievable sum rate of all users.
We can observe that all four curves ascend as $K$ increases within a certain range. This is because the more users are in the network, the higher utilization of network resources is obtained. Further, such phenomenon also results in greater total network benefit (i.e., sum rate of all users). Due to limited network resources, the sum rate cannot be infinitely improved by increasing $K$.
Moreover, benefiting from the optimal gain brought by the PBF at IRS and the effective FRA-based UA algorithm, our proposed algorithm achieves optimal performance especially in high load conditions.
The simulation result in Fig. \ref{K} presents that when $K=48$, our proposed algorithm can improve the performance compared with AF relay algorithm, UAPC algorithm, NBUA algorithm and benchmark by $15\%$, $24\%$, $39\%$ and $70\%$.

Next, we fix $K=16$, $M=32$, $N=32$ and plot the cumulative distribution function (CDF) of sum rate as shown in Fig. \ref{CDF}.
It can be seen that, the performance gains of the four algorithms are stable, and also keep consistent with their counterparts in Fig. \ref{M} and Fig. \ref{K}. 
We can observe that our proposed algorithm has a probability of about $0.92$ to obtain achievable rate $R_{sum}>115.7$ bps/Hz (i.e., $R_{k}>7.23$ bps/Hz).
Therefore, we conclude that the performance of our proposed algorithm performs the best and can adapt to most of scenarios.  

\subsection{Coverage Performance and Average Rate Performance per User}
\begin{figure}
	[t]
	\centering
	\captionsetup{font={small}}
	\captionsetup{singlelinecheck = false, justification=justified}
	\begin{minipage}[t]{0.452\linewidth}
		\centerline{\includegraphics[angle=0, width=2.84in,height=2.3in]{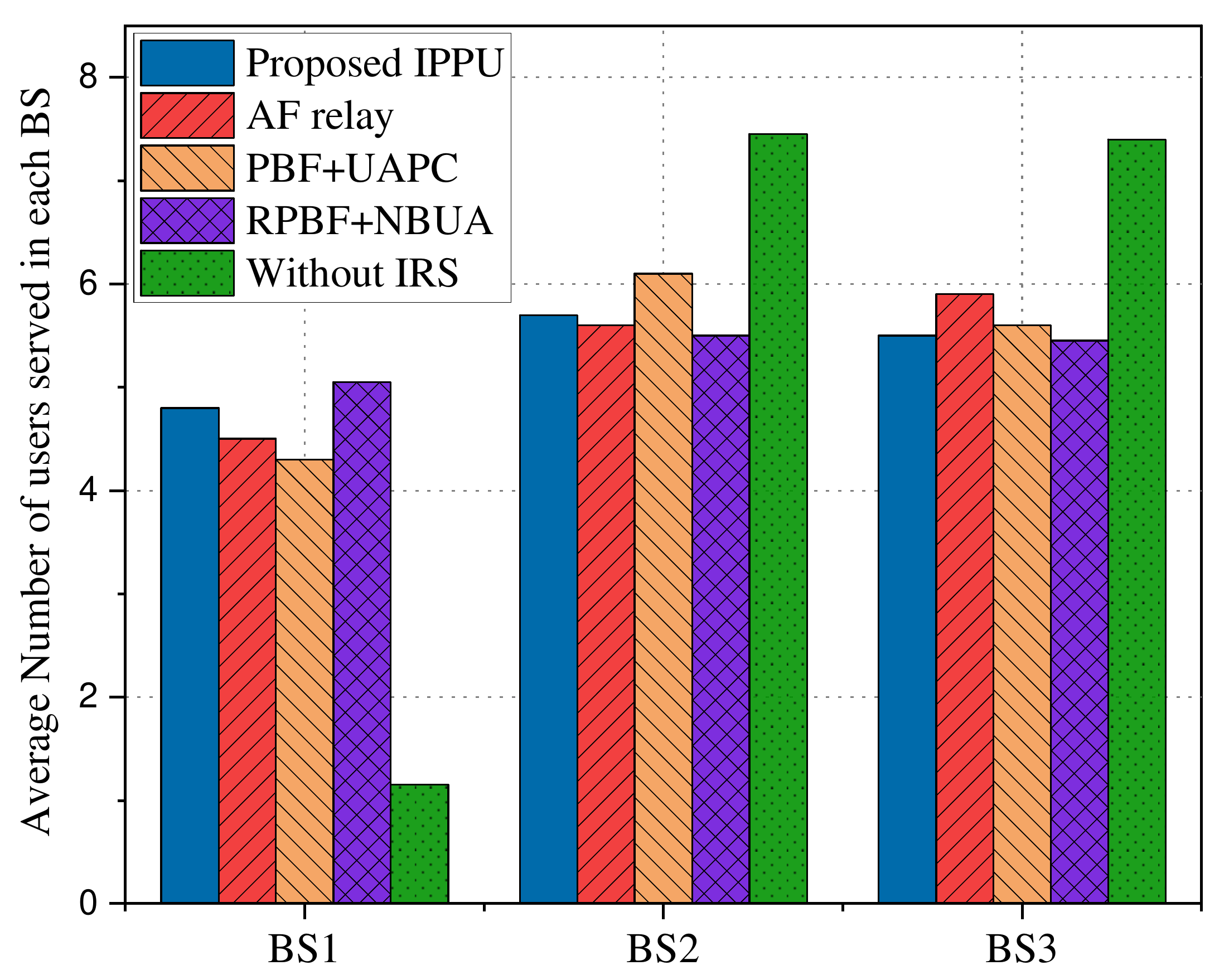}}
		\caption{Average number of users per BS served for four comparison algorithms with $K=16$, $M=32$ and $N=32$.}
		\label{Cover}
	\end{minipage}  \hspace{2.0em}
	\begin{minipage}[t]{0.452\linewidth}
		\centerline{\includegraphics[angle=0, width=2.88in,height=2.32in]{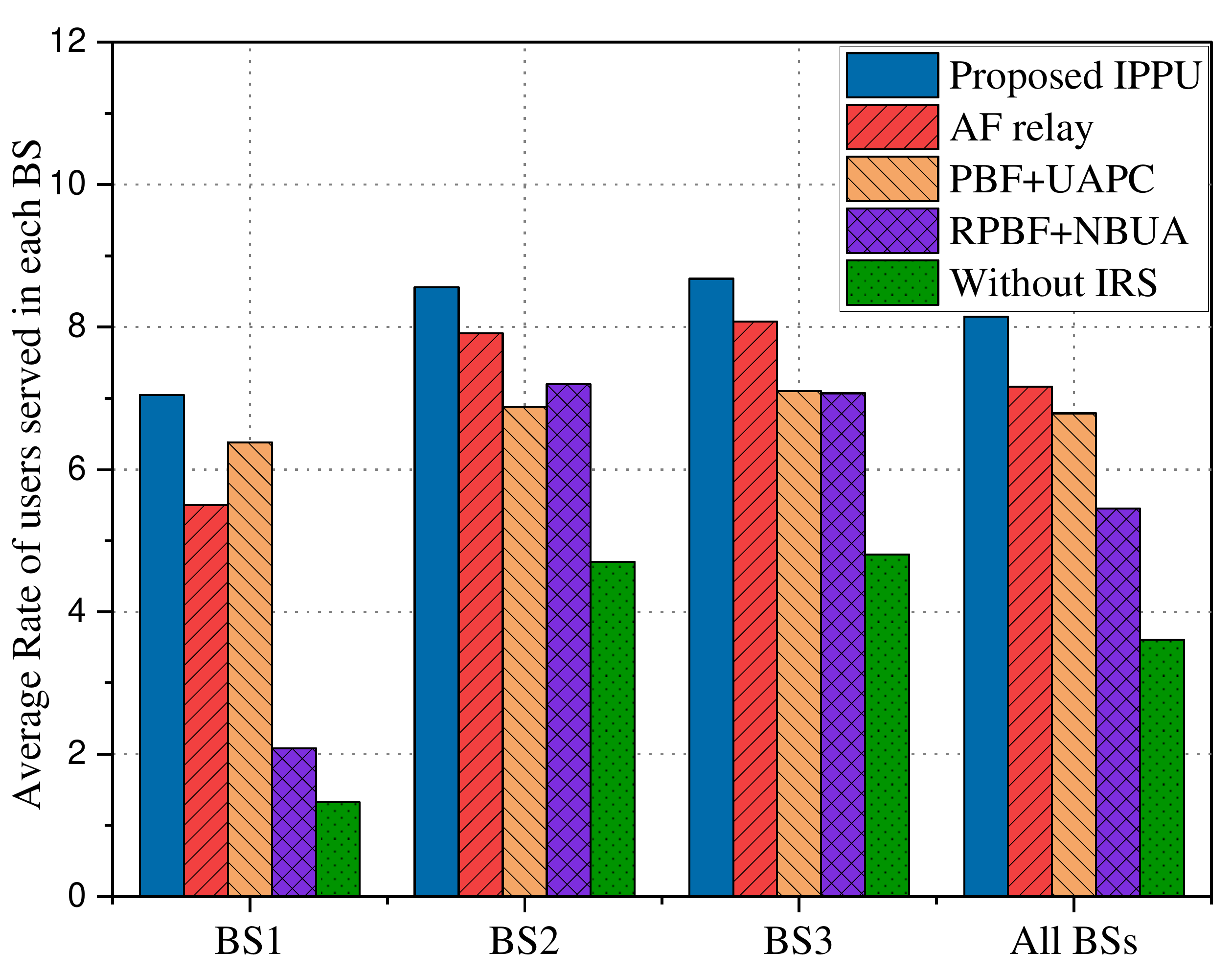}}
		\caption{Average rate of users per BS served for four comparison algorithms with $K=16$, $M=32$ and $N=32$.}
		\label{Cover_R}
	\end{minipage}	
	\vspace{-0.5em}
\end{figure}

Fig. \ref{Cover} compares the average total number of users served by each BS of different algorithms, where we average the results of $10^3$ scenarios and the same as below.
Observing the number of users served by heavily blocked BS 1, it is obvious that our algorithm serves the most users compared with the other three algorithms except NBUA (UA is only related to physical distance) algorithm, and the benchmark without IRS serves the smallest number of users. This demonstrates that IRS can expand the coverage of mmWave signal and increase the number of users served by the blocked BS.
In addition, the number of users served by BS 2 and BS 3 are relatively close under different algorithms, which is consistent with the actual situation.

\begin{figure}
	[t]
	\captionsetup{font={small}}
	\captionsetup{singlelinecheck = false, justification=justified}
	\centering
	\includegraphics[width=0.5\columnwidth]{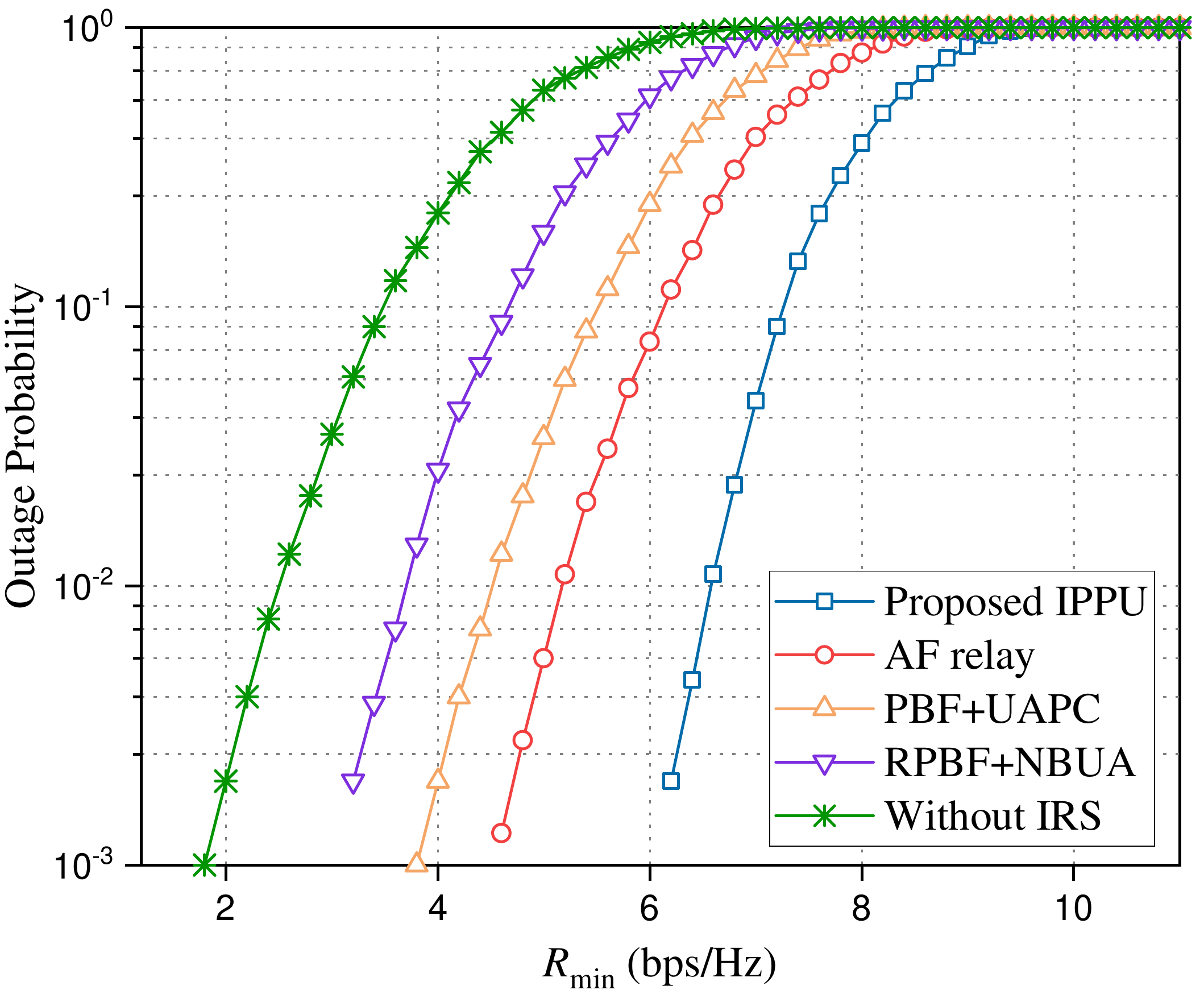}
	\caption{Outage probability versus $R_{min}$ with $K=16$, $M=32$ and $N=32$.}
	\label{Out_P}
	\vspace{-1.5em}
\end{figure}

Fig. \ref{Cover_R} shows the average rate of users served by each BS.
It can be seen that due to the optimization of the PBF at the IRS, both our algorithm and the UAPC algorithm can provide a higher communication rate for users served by BS 1 and AF relay also achieves a good average rate gain for similar reasons. In contrast, the remaining two algorithms result in lower average rates due to little or no PBF gain (i.e., the communication between users and BS 1 is greatly affected by obstacles). For the entire mmWave system, the relationship of average rate of users also corresponds to the results in previous subsections IV-C and IV-D.
\vspace{-0.5em}
\subsection{Outage Probability Performance versus $R_{min}$}
Finally, in order to investigate the impact of different minimum rate constraints on all algorithms and show the robustness of our IRS-assisted system against blockages, we calculate the outage probability as follows
\begin{equation}\label{outP}
P_{out}(R_{min})=\operatorname{Pr}\left(E[R_k]\leq R_{min}\right),
\end{equation}
where $\operatorname{Pr}(\cdot)$ is the probability function. 
From Fig. \ref{Out_P}, we observe that the outage probability can be substantially reduced by deploying IRS and exploiting our IPPU algorithm. Under the same minimum rate constraints, our proposed algorithm has the lowest outage probability of communication. 
Also, as $R_{min}$ increases, the outage probability of user communication continues to rise. Especially, when $R_{min}=8.4$ bps/Hz, the outage probability of these algorithms are close to $1$ except our IPPU algorithm.

\section{Conclusion}
In this paper, we design an IRS-assisted multi-BS multi-user mmWave system to realize high robust, cost-effective and high coverage mmWave communication. 
With consideration of the impact of IRS on UA when multiple BSs are deployed, a maximization sum rate problem is formulated by jointly optimizing PBF at IRS with discrete phase constraint, power allocation with limited transmit power and UA with QoS constraints. Then, an efficient iterative algorithm based on alternate optimization is proposed to solve this non-convex problem. 
Extensive simulation results corroborate the feasibility and effectiveness of our proposed algorithm under various scenarios.
Specifically, our proposed algorithm converges at an acceptable rate and obtains significant sum rate promotion.
In particular, our proposed algorithm is able to provide up to $175\%$ higher sum rate compared with the benchmark and $140\%$ higher EE compared with AF relay.

In the future, it will be desirable to investigate multi-IRS assisted multi-BS mmWave communication with power allocation and UA. The collaborative work of multi-IRS can provide more optimization variables to
tune, thereby further enhancing the performance of the system.
Other meaningful research directions include the acquisition of CSI and the impact of channel estimation on the performance of IRS-assisted systems. 
Channel estimation at IRS is meaningful and important as we have mentioned, which will be more complicated in the IRS-assisted mmWave systems by jointly considering with power allocation and UA. 
And it is worthy to find a suitable channel estimation method for higher estimation accuracy with lower feedback overhead.
\appendices
\section{Proof of Theorem 1}
Rewrite the power allocation matrix $\mathbf{P}_i$ as $\mathbf{P}_i=\mathbf{Q}\mathbf{Q}^T$ with $\mathbf{Q}=\sqrt{\mathbf{P}_i}$, and let $\mathbf{\widetilde{H}}^i_r=\mathbf{Q}^{-1}\mathbf{H}^i_{r}$. Then the objective function $f_1(\mathbf{\Phi})$ in (\ref{p3}) can be rewritten as
\begin{equation}\label{f1}
\begin{split}
f_1(\mathbf{\Phi}) &=\operatorname{tr}((\mathbf{Q}^{-1}\mathbf{H}^i_{r}\mathbf{\Phi}\mathbf{G}^i)^+ (\mathbf{Q}^{-1} \mathbf{H}^i_{r}\mathbf{\Phi}\mathbf{G}^i)^{+H}) \\
& =\operatorname{tr}((\mathbf{\widetilde{H}}^i_r\mathbf{\Phi}\mathbf{G}^i)^+ (\mathbf{\widetilde{H}}^i_r\mathbf{\Phi}\mathbf{G}^i)^{+H}) \\
&\overset{a}{=}\|(\mathbf{G}^i)^+\mathbf{\Phi}^{-1}(\mathbf{\widetilde{H}}^i_r)^+\|^2_F\\
&\overset{b}{=}\|\operatorname{vec}((\mathbf{G}^i)^+\mathbf{\Phi}^{-1}(\mathbf{\widetilde{H}}^i_r)^+)\|^2\\
&=\|(\mathbf{\widetilde{H}}^i_r)^{+H}\otimes (\mathbf{G}^i)^+\operatorname{vec}(\mathbf{\Phi}^{-1})\|^2\\
&=\operatorname{vec}(\mathbf{\Phi}^{-1})^H ((\mathbf{\widetilde{H}}^i_r)^{+H}\otimes (\mathbf{G}^i)^+)^H ((\mathbf{\widetilde{H}}^i_r)^{+H}\otimes (\mathbf{G}^i)^+) \operatorname{vec}(\mathbf{\Phi}^{-1})
\end{split}
\end{equation}
Here, we have applied the properties of the Frobenius matrix norm and the pseudo-inverse law of matrix product in step (a), whereas steps (b) follows from the vectorization operator.
Thus, the optimization problem (\ref{13}) is equivalent to (\ref{f1}).

%
\ifCLASSOPTIONcaptionsoff
\newpage
\fi

\end{document}